\documentclass[journal]{IEEEtran}

\pdfoutput=1

\usepackage{amssymb,amsfonts,amsmath,amscd,dsfont,mathrsfs}
\usepackage{amsthm}
\usepackage[utf8]{inputenc}

\usepackage{booktabs} 
\usepackage{graphicx,tikz,standalone,rotating}
\usepackage{algorithm}
\usepackage{algorithmic}
\usepackage{flushend}

\usepackage{url}

\usepackage{fakufakumath}

\newcommand{\whtpair}{\overset{\text{WHT}}{\longleftrightarrow}}
\newcommand{\GL}{\operatorname{GL}(n,\F_2)}
\newcommand{\mN}{\mathbb{N}}
\newcommand{\var}{\text{Var}}
\newcommand{\Hash}{\mathcal{H}}

\providecommand{\G}{{\cal G}}

\graphicspath{{./figures/}}

\title{
  A Fast Hadamard Transform for Signals with Sub-linear Sparsity in the Transform Domain
}
\author{Robin~Scheibler,~\IEEEmembership{Student Member,~IEEE}
  Saeid~Haghighatshoar,~\IEEEmembership{Student Member,~IEEE}
  Martin~Vetterli,~\IEEEmembership{Fellow,~IEEE}%
  \thanks{R. Scheibler, S. Haghighatshoar and Martin Vetterli are with
    the School of Computer and Communication Sciences \'Ecole Polytechnique
    F\'ed\'erale de Lausanne (EPFL), CH-1015 Lausanne, Switzerland.}%
  \thanks{Email: \{robin.scheibler, saeid.haghighatshoar, martin.vetterli\}@epfl.ch}%
  \thanks{The research of Robin Scheibler was supported by ERC Advanced
    Investigators Grant: Sparse Sampling: Theory, Algorithms and Applications
    SPARSAM no. $247006$.}%
  \thanks{A short version of this paper was presented at the 51st Annual
    Allerton Conference on Communication, Control, and Computing, Monticello, 2013.}%
}

\begin{document}

  \maketitle

  \begin{abstract}
    A new iterative low complexity algorithm has been presented for computing the
    Walsh-Hadamard transform (WHT) of an $N$ dimensional signal with a $K$-sparse
    WHT, where $N$ is a power of two and $K=O(N^\alpha)$, scales sub-linearly in
    $N$ for some $0<\alpha<1$.  Assuming a random support model for the non-zero
    transform domain components, the algorithm reconstructs the WHT of the signal
    with a sample complexity $\bigO(K \log_2(\frac{N}{K}))$, a computational
    complexity $\bigO(K\log_2(K) \log_2(\frac{N}{K}))$ and with a very high
    probability asymptotically tending to $1$. 

    The approach is based on the subsampling (aliasing) property of the WHT,
    where by a carefully designed subsampling of the time domain signal, one can
    induce a suitable aliasing pattern in the transform domain. By treating  the
    aliasing patterns as parity-check constraints and borrowing ideas from
    erasure correcting sparse-graph codes, the recovery of the non-zero spectral
    values has been formulated as a belief propagation (BP) algorithm (peeling
    decoding) over a sparse-graph code for the binary erasure channel (BEC).
    Tools from coding theory are used to analyze the asymptotic performance of
    the algorithm in the \textit{``very sparse''} ($\alpha \in (0,\frac{1}{3}]$)
    and the \textit{``less sparse''} ($\alpha \in (\frac{1}{3}, 1)$) regime. 
  \end{abstract}


  \section{Introduction}

  \IEEEPARstart{T}{he} fast Walsh-Hadamard transform (WHT) is a well-known
  signal processing tool with application in areas as varied as image
  compression and coding \cite{Pratt:1969hq}, spreading sequence for multi-user
  transmission in cellular networks (CDMA) \cite{3gpp_ts25213}, spectroscopy
  \cite{Horadam} as well as compressed sensing \cite{saeid}. It has also nice
  properties studied in different areas of mathematics \cite{hedayat}.  Its
  recursive structure, similar to the famous fast Fourier transform (FFT)
  algorithm for computing the discrete Fourier transform (DFT) of the signal,
  allows a fast computation with complexity $\bigO(N \log_2(N))$ in the
  dimension of the signal $N$ \cite{Lee:1986bt,Johnson:2000bo}.

  A number of recent publications have addressed the particular problem of
  computing the DFT of an $N$ dimensional signal under the assumption of
  $K$-sparsity of the signal in the frequency domain
  \cite{Gilbert:2002wu,Gilbert:2008ba,Lawlor:2012wc,Hassanieh:2012uq,Hassanieh:2012wd}.
  In particular, it has been shown that the well known computational complexity
  $\bigO(N\log_2(N))$ of the FFT algorithm can be strictly improved.  Such
  algorithms are generally known as \textit{sparse} FFT (sFFT) algorithms.  The
  authors in \cite{Ghazi:2013vs} by extending the  results of
  \cite{Hassanieh:2012wd}, gave a very low complexity algorithm for computing
  the 2D-DFT of a $\sqrt{N}\times\sqrt{N}$ signal.  In a similar line of work,
  based on the subsampling property of the DFT in the time domain resulting in
  aliasing in the frequency domain, the authors in
  \cite{Pawar:2012df,Pawar:2013vm} developed a novel low complexity iterative
  algorithm to recover the non-zero frequency elements using ideas from
  sparse-graph codes \cite{richardson}. 

  In this paper, we develop a fast algorithm to compute the WHT
  of data sparse in the Hadamard domain. We first develop some useful
  properties of the WHT, specially the subsampling and the modulation property,
  that will later play a vital role in the development the algorithm. In
  particular, we show the subsampling in time domain allows to induce a
  well-designed aliasing pattern over the transform domain components. In other
  words, it is possible to obtain a  linear combination of a controlled
  collection of transform domain components (aliasing), which creates
  interference between the non-zero components if more than one of them are
  involved in the induced linear combination. Similar to \cite{Pawar:2013vm}
  and borrowing ideas from sparse-graph codes, we construct a bipartite graph
  by considering the non-zero values in the transform domain as variable nodes
  and interpreting any induced aliasing pattern as a parity check constraint
  over the variables in the graph.  We analyze the structure of the resulting
  graph assuming a random support model for the non-zero coefficients in the
  transform domain. Moreover, we give an iterative peeling decoder to recover
  those non-zero components. In a nutshell, our proposed sparse fast Hadamard
  transform (SparseFHT) consists of a set of deterministic linear hash
  functions (explicitly constructed) and an iterative peeling decoder that uses
  the hash outputs to recover the non-zero transform domain variables.  It
  recovers the $K$-sparse WHT of the signal in sample complexity (number of
  time domain samples used) $\bigO(K \log_2(\frac{N}{K}))$, total computational
  complexity $\bigO(K \log_2(K) \log_2(\frac{N}{K}))$ and with a high
  probability approaching $1$ asymptotically, for any value of $K$.

  \vspace{.5mm}
  {\bf Notations and Preliminaries:} For $m$ an integer, the set of all integers
  $\{0,1,\dots,m-1\}$ is denoted by $[m]$.  We use the small letter $x$ for the
  time domain and the capital letter $X$ for the transform domain signal. For an
  $N$ dimensional real-valued vector $v$, with $N=2^n$ a power of two, the $i$-th
  components of $v$ is equivalently represented by $v_i$ or
  $v_{i_0,i_1,\ldots,i_{n-1}}$, where $i_0,i_1,\dots, i_{n-1}$ denotes the binary
  expansion of $i$ with $i_0$ and $i_{n-1}$ being the least and the most
  significant bits. Also sometimes the real value assigned to $v_i$ is not
  important for us and by $v_i$ we simply mean the binary expansion associated to
  its index $i$, however, the distinction must be clear from the context. $\F_2$
  denotes the binary field consisting of $\{0,1\}$ with summation and
  multiplication modulo $2$. We also denote by $\F_2^n$ the space of all $n$
  dimensional vectors with binary components and  addition of the vectors
  done component wise. The inner product of two $n$ dimensional binary vectors
  $u,v$ is defined by $\ip{u}{v}=\sum_{t=0}^{n-1} u_t v_t$ with arithmetic over
  $\F_2$ although $\ip{}{}$ is not an inner product in exact mathematical sense,
  for example, $\ip{u}{u}=0$ for any $u \in \F_2^n$.

  For a signal $X \in \R^N$, the support of $X$ is defined as $\supp(X)=\{i \in
    [N]: X_i \neq 0\}$. The signal $X$ is called $K$-sparse if $|\supp(X)|=K$,
  where for a set $A\subset [N]$, $|A|$ denotes the cardinality or the number of
  elements of $A$. For a collection of $N$ dimensional signals ${\cal S}_N
  \subset \R^N$, the sparsity of ${\cal S}_N$ is defined as $K_N=\max _{X \in
    {\cal S}_N} |\supp(X)|$. 

  \begin{definition}
    A class of signals  of increasing dimension $\{{\cal S}_N\}_{N=1}^\infty$ has
    sub-linear sparsity if there is $0<\alpha<1$ and some $N_0 \in \N$ such that
    for all $N>N_0$, $K_N \leq N^\alpha$. The value $\alpha$ is called the sparsity
    index of the class.
  \end{definition}

  \section{Main Results}

  Let us first describe the main result of this work in the following theorem.

  \begin{theorem}\label{main_theorem}
    Let $0<\alpha<1$, $N=2^n$ a power of two and $K=N^\alpha$. Let $x \in \R^N$ be
    a time domain signal with a WHT $X \in \R^N$. Assume that $X$ is a $K$-sparse
    signal in a class of signals with sparsity index $\alpha$ whose support is
    uniformly at random selected among all possible ${N \choose K}$ subsets of $[N]$
    of size $K$. For any value of $\alpha$, there is an algorithm that can compute
    $X$ and has the following properties:
    \begin{enumerate}
      \item {\bf Sample complexity:} The algorithm uses $C K
            \log_2(\frac{N}{K})$ time domain samples of the signal $x$. $C$ is
            a function of $\alpha$ and $C \leq (\frac{1}{\alpha} \vee
            \frac{1}{1-\alpha})+1$, where for $a,b \in \R_+$, $a \vee b$
            denotes the maximum of $a$ and $b$.

      \item {\bf Computational complexity:} The total number of operations in
            order to successfully decode all the non-zero spectral components
            or announce a decoding failure is $\bigO(C K \log_2(K)
            \log_2(\frac{N}{K}))$.

      \item {\bf Success probability:} The algorithm correctly computes the
            $K$-sparse WHT $X$ with very high probability asymptotically
            approaching $1$ as $N$ tends to infinity, where the probability is
            taken over all random selections of the support of $X$.
    \end{enumerate}
  \end{theorem}

  \begin{remark}
    To prove Theorem \ref{main_theorem}, we distinguish between the very sparse
    case ($0 < \alpha \leq \frac{1}{3}$) and less sparse one ($\frac{1}{3} <
    \alpha <1$). Also, we implicitly assume that the algorithm knows the value
    of $\alpha$ which might not be possible in some cases. As we will see later
    if we know to which regime the signal belongs and some bounds on the value
    of the $\alpha$, it is possible to design an algorithm that works for all
    those values of $\alpha$. However, the sample and computational complexity
    of that algorithm might increase compared with the optimal one that knows
    the value of $\alpha$. For example, if we know that the signal is very
    sparse, $\alpha \in (0, \alpha^*]$ with $\alpha^* \leq \frac{1}{3}$, it is
    sufficient to design the algorithm for $\alpha^*$ and it will work for all
    signals with sparsity index less that $\alpha^*$. Similarly, if the signal
    is less sparse with a sparsity index  $\alpha \in (\frac{1}{3}, \alpha^*)$,
    where $\alpha^* <1$, then again it is sufficient to design the algorithm
    for $\alpha^*$ and it will automatically work for all $\alpha \in
    (\frac{1}{3}, \alpha^*)$.
  \end{remark}

  \begin{remark}\label{remark_verysparse}
    In the very sparse regime ($0<\alpha \leq \frac{1}{3}$), we prove that for
    any value of $\alpha$ the success probability of the optimally designed
    algorithm is at least $1- O(1/K^{3 (C/2 \, -1)})$, with
    $C=[\frac{1}{\alpha}]$ where for $u\in \R_+$, $[u]=\max \{n \in \Z: n \leq
      u\}$. It is easy to show that for every value of $\alpha \in (0,
    \frac{1}{3})$ the success probability  can be lower bounded by
    $1-O(N^{-\frac{3}{8}})$.
  \end{remark}


  \section{Walsh-Hadamard Transform and its Properties}

  Let $x$ be an $N=2^n$ dimensional signal indexed with elements $m \in
  \F_2^n$. The $N$ dimensional WHT of  the signal $x$ is defined by $$
  X_k=\frac{1}{\sqrt{N}} \sum _{m \in \F_2^n} (-1)^{\ip{k}{m}} x_m,$$ where $k
  \in \F_2^n$ denote the corresponding binary index of the transform domain
  component. Also, throughout the paper, borrowing some terminology from the
  DFT, we call transform domain samples $X_k, k\in \F_2^n$ frequency or
  spectral domain components of the time domain signal $x$.

  %

  \subsection{Basic Properties}

  This subsection is devoted to reviewing some of the basic properties of the
  WHT. Some of the properties are not directly used in the paper and we have
  included them for the sake of completeness. They can be of independent
  interest. The proofs of all the properties are provided in \aref{prop_proof}.

  \begin{property}[Shift/Modulation]
    \label{had_shift}
    Let $X_k$ be the WHT of the signal $x_m$ and let $p\in\F_2^n$. Then
    \begin{align*}
      {x_{m + p} \quad \whtpair \quad X_k (-1)^{\ip{p}{k}}}.
    \end{align*}
  \end{property}

  The next property is more subtle and allows to partially permute the Hadamard
  spectrum in a specific way by applying a corresponding permutation in the
  time domain. However, the collection of all such possible permutations is
  limited. We give a full characterization of all those permutations.
  Technically, this property is equivalent to finding permutations $\pi_1,
  \pi_2:[N] \to[N]$ with corresponding permutation matrices $\Pi_1,\Pi_2$ such that
  \begin{equation}
    \Pi_2 H_N = H_N \Pi_1,
    \label{perm_prop}
  \end{equation}
  where $H_N$ is the Hadamard matrix of order $N$ and where the permutation
  matrix corresponding to a permutation $\pi$ is defined by $(\Pi)_{i,j}=1$ if
  and only if $\pi(i)=j$, and zero otherwise.  The identity \eqref{perm_prop}
  is equivalent to finding a row permutation of $H_N$ that can be equivalently
  obtained by a column permutation of $H_N$.
  \begin{property}
    \label{gln2perm}
    All of the permutations satisfying \eqref{perm_prop} are described by the elements of 
    \begin{equation}
      \GL = \{A \in \F_2^{n\times n} \,\vert\, A^{-1}\  \text{exists}\},
      \nonumber
    \end{equation}
    the set of $n\times n$ non-singular matrices with entries in $\F_2$.
  \end{property}

  \begin{remark}
    The total number of possible permutations in  Property \ref{gln2perm}, is
    $\prod_{i=0}^{n-1} (N-2^i)$, which is a negligible fraction of all $N!$
    permutation over $[N]$.
  \end{remark}

  \begin{property}[Permutation]
    \label{had_perm}
    Let $\Sigma \in \GL$. Assume that $X_k$ is the WHT of the time domain
    signal $x_m$. Then
    \begin{equation}
      x_{\Sigma m} \quad \whtpair \quad X_{\Sigma^{-T}k}.
      \nonumber
    \end{equation}
  \end{property}

  \begin{remark}
    Notice that any $\Sigma \in \GL$ is a bijection from $\F_2^n$ to $\F_2^n$,
    thus $x_{\Sigma m}$ is simply a vector obtained by permuting the initial
    vector $x_m$.
  \end{remark}

  The last property is that of downsampling/aliasing. Notice that for a vector
  $x$ of dimension $N=2^n$, we index every components by a binary vector of
  length $n$, namely, $x_{m_0,m_1,\dots,m_{n-1}}$. To subsample this vector
  along dimension $i$, we freeze the $i$-th component of the index to either
  $0$ or $1$. For example, $x_{0,m_1,\dots,m_{n-1}}$ is a $2^{n-1}$ dimensional
  vector obtained by subsampling the vector $x_m$ along the first index. 

  \begin{property}[Downsampling/Aliasing]
    \label{had_ds}
    Suppose that $x$ is a vector of dimension $N=2^n$ indexed by the elements
    of $\F_2^n$ and assume that $B=2^b$, where $b\in \mN$ and $b<n$. Let   
    \begin{equation}
      \Psi_b = \left[ \, \mathbf{0}_{b\times (n-b)} \ I_b \, \right]^T,
      \elabel{ds_mat}
    \end{equation}
    be the subsampling matrix freezing the first $n-b$ components in the index 
    to $0$. If $X_k$ is the WHT of $x$, then
    \begin{equation}
      x_{\Psi_b m} \quad \whtpair \quad \sqrt{\frac{B}{N}} \sum_{j\in\Null\left(\Psi_b^T\right)} X_{\Psi_b k+ j},
      \nonumber
    \end{equation}
    where $x_{\Psi_b m}$ is a $B$ dimensional signal labelled with $m \in \F_2^b$.
  \end{property}
  Notice that Property \ref{had_ds} can be simply applied for any matrix
  $\Psi_b$ that subsamples any set of indices of length $b$ not necessarily the
  $b$ last ones. 
  \begin{remark}
    The group $\F_2^n$ can be visualized as the vertices of the $n$-dimensional
    hypercube. The downsampling property just explained implies that downsampling
    along some of the dimensions in the time domain is equivalent to summing up all
    of the spectral components along \textit{the same dimensions} in the spectral
    domain.  This is illustrated visually in \ffref{cube_ds} for dimension $n=3$.
  \end{remark}

  \begin{figure}
    \centering
    \includegraphics[width=\linewidth]{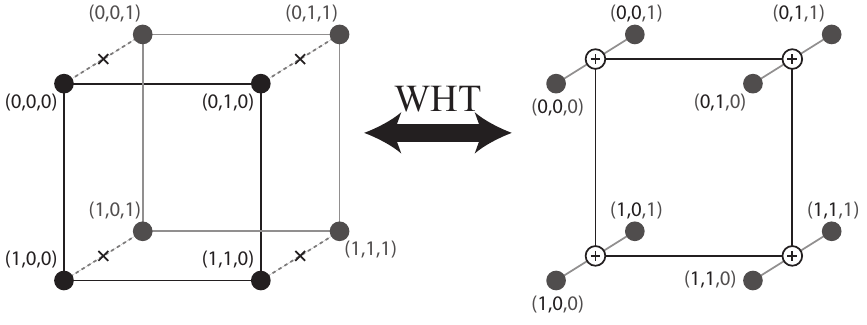}
    \caption{Illustration of the downsampling property on a hypercube for
      $N=2^3$. The two cubes are the time-domain and Hadamard-domain signals on
      the left and right, respectively.  We decide to drop all nodes whose
      third coordinate is '1'. We illustrate this by adding an '$\times$' on
      the edges connecting these vertices through the third coordinate.  This
      is equivalent to summing up vertices along the corresponding edges in the
      Hadamard domain.
    }
    \flabel{cube_ds}
  \end{figure}

  \begin{remark}
    In a general downsampling procedure, one can replace  the frozen indices by
    an arbitrary but fixed binary pattern. The only difference is that instead
    of summing the aliased spectral components, one should also take into
    account the suitable $\{+,-\}$ sign patterns, namely, we have 
    \begin{equation}
      x_{\Psi_b m +p} \quad \whtpair \quad \sqrt{\frac{B}{N}} \sum_{j\in\Null\left(\Psi_b^T\right)} (-1)^{\ip{p}{j}} X_{\Psi_b k+ j},\label{p_pattern}
    \end{equation}
    where $p$ is a binary vector of length $n$ with  $b$ zeros at the end. 
  \end{remark}

  \section{Hadamard Hashing Algorithm}
  \slabel{hashing}

  By applying the basic properties of the WHT, one can design suitable hash
  functions in the spectral domain. The main idea is that one does not need to
  have access to the spectral values and the output of all hash functions can
  be simply computed by low complexity operations on the time domain samples of
  the signal.  

  \begin{proposition}[Hashing]
    \plabel{hashing}
    Assume that $\Sigma\in\GL$ and $p\in\F_2^n$. Let $N=2^n$, $b \in \mN$,
    $B=2^b$ and let $m,k\in\F_2^b$ denote the time and frequency indices of a
    $B$ dimensional signal and its WHT defined by 
    \begin{equation*}
      u_{\Sigma,p}(m) = \sqrt{\frac{N}{B}}\,x_{\Sigma\Psi_b m + p}.  
    \end{equation*}
    Then, the length $B$ Hadamard transform of $u_{\Sigma,p}$ is given by
    \begin{equation}
      U_{\Sigma,p}(k) = \sum_{j \in \F_2^n\,|\,\Hash j=k} X_j\, (-1)^{\ip{p}{j}},
      \elabel{hash_bucket}
    \end{equation}
    where $\Hash$ is the index hashing operator defined by 
    \begin{equation}
      \Hash = \Psi_b^T\Sigma^T,
      \elabel{hashing}
    \end{equation}
    where $\Psi_b$ is as in \eref{ds_mat}. Note that the index of components in
    the sum \eref{hash_bucket} can be explicitely written as function of the bin index $k$
    \begin{align*}
      j=\Sigma^{-T}\Psi_bk+q,\qquad q\in\Null(\Hash).
    \end{align*}
  \end{proposition}
  The proof simply follows from the properties \ref{had_shift}, \ref{had_perm}, and
  \ref{had_ds}.

  Based on \pref{hashing}, we give \algref{fasthadhash} which
  computes the hashed Hadamard spectrum. Given an FFT-like fast Hadamard
  transform (FHT) algorithm, and picking $B$ bins for hashing the spectrum,
  \algref{fasthadhash} requires $\bigO(B\log B)$ operations.

  \begin{algorithm}
    \caption{FastHadamardHashing$(x, N, \Sigma, p, B)$}
    \alabel{fasthadhash}
    \begin{algorithmic}
      \REQUIRE Signal $x$ of dimension $N=2^n$, $\Sigma$ and $p$ and given number of output bins $B=2^b$ in a hash.
      \ENSURE $U$ contains the hashed Hadamard spectrum of $x$.
      \STATE $u_m = x_{\Sigma\Psi_b m+p}$, for $m\in\F_2^b$.
      \vspace{0.5mm}
      \STATE $U = \sqrt{\frac{N}{B}}\operatorname{FastHadamard}(u_m, B)$.
    \end{algorithmic}
  \end{algorithm}

  \subsection{Properties of Hadamard Hashing}

  In this part, we review some of the nice properties of the hashing algorithm
  that are crucial for developing an iterative peeling decoding algorithm to
  recover the non-zero spectral values.  We explain how it is possible to
  identify collisions between the non-zero spectral coefficients that are hashed
  to the same bin and also to estimate the support of non-colliding components. 

  Let us consider $U_{\Sigma,p}(k)$ for two cases: $p=0$ and some $p\neq 0$. It
  is easy to see that in the former $U_{\Sigma,p}(k)$ is obtained by summing
  all of the spectral variables hashed to bin $k$ -- those whose index $j$ satisfies
  $\Hash j=k$ -- whereas in the latter the same variables are added together
  weighted by $(-1)^{\ip{p}{j}}$. Let us define the following ratio test
  \begin{equation}
    r_{\Sigma,p}(k) = \frac{U_{\Sigma,p}(k)}{U_{\Sigma,0}(k)}.
    \nonumber
  \end{equation}
  When the sum in $U_{\Sigma,p}(k)$ contains only one non-zero component, it is
  easy to see that $|r_{\Sigma,p}(k)|=1$ for \textit{`any value'} of $p$.
  However, if there is more than one component in the sum, under a very mild
  assumption on the the non-zero coefficients of the spectrum (i.e. they are
  jointly sampled from a continuous distribution), one can show that
  $|r_{\Sigma,p}(k)|\neq 1$ for at least some values of $p$. In fact, $n-b$
  well-chosen values of $p$ allow to detect whether there is only one, or more
  than one non-zero components in the sum.

  When there is only one $X_{j^\prime}\neq0$ hashed to the bin $k$
  ($h_{\Sigma}(j^\prime)=k$), the result of the ratio test is precisely $1$ or
  $-1$, depending on the value of the inner product between $j^\prime$ and $p$.
  In particular, we have 
  \begin{equation}
    \ip{p}{j^\prime} = \indicator{r_{\Sigma,p}(k) < 0},
    \elabel{ip_est}
  \end{equation}
  where $\indicator{t<0} = 1$ if $t<0$, and zero otherwise.  Hence, if for
  $n-b$ well-chosen values of $p$, the ratio test results in $1$ or $-1$,
  implying that there is only one non-zero spectral coefficient in the
  corresponding hash bin, by some extra effort it is even possible to identify
  the position of this non-zero component. We formalize this result in the
  following proposition proved in \aref{sup_est_proof}.

  \begin{proposition}[Collision detection / Support estimation]
    \plabel{collision_detect}
    Let $\Sigma \in\GL$ and let $\sigma_i, i\in [n]$ denote the columns of $\Sigma$.
    \begin{enumerate}
      \item If for all $d\in[n-b]$, $|r_{\Sigma,\sigma_d}(k)|=1$ then almost
        surely there is  only one non-zero spectral value in the bin
        indexed by $k$. Moreover, if we define
        \begin{equation}
          \hat{v}_d = \begin{cases}
            \indicator{r_{\Sigma,\sigma_d}(k) < 0} & d \in [n-b],\\
            0 & \text{otherwise,}
          \end{cases}
          \nonumber
        \end{equation}
        the index of the unique non-zero coefficient is given by 
        \begin{equation}
          j = \Sigma^{-T}(\Psi_b\,k + \hat{v}).
          \elabel{index_rec}
        \end{equation}
      \item If there exists a $d \in[n-b]$ such that $|r_{\Sigma,\sigma_d}(k)| \neq
        1$ then the bin $k$ contains more than one non-zero coefficient.
    \end{enumerate}
  \end{proposition}

  \begin{figure*}
    \centering
    \includegraphics[width=0.8\linewidth]{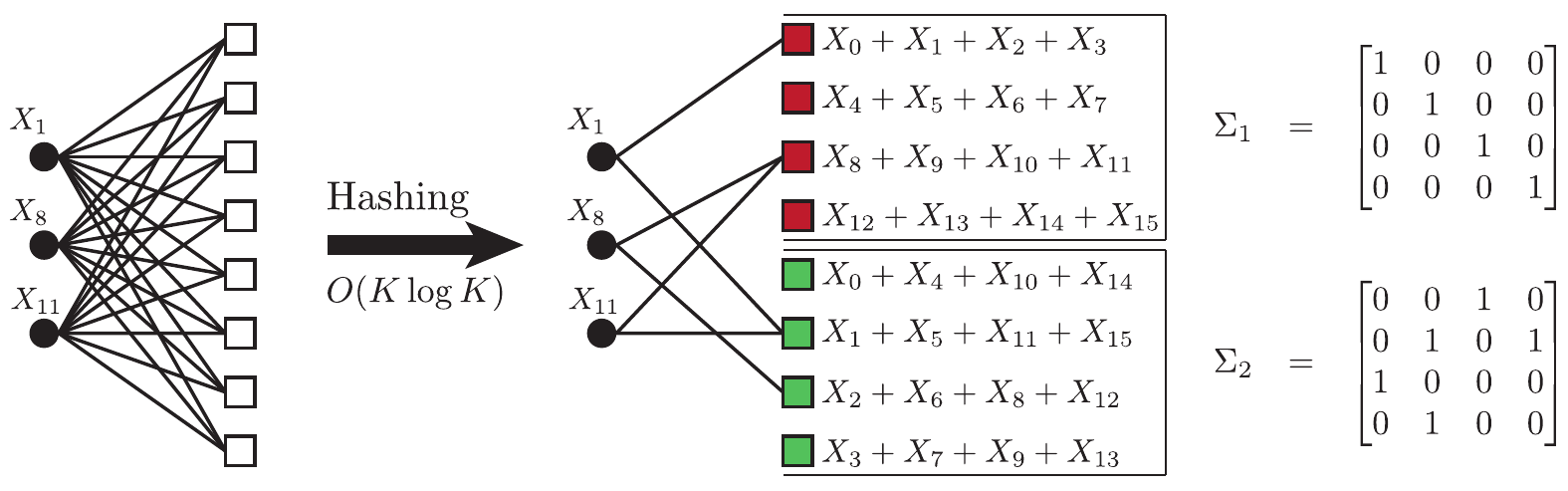}
    \caption{On the left, bipartite graph representation of the WHT for $N=8$
      and $K=3$.  On the right, the underlying bipartite graph after applying
      $C=2$ different hashing produced by plugging $\Sigma_1$, $\Sigma_2$ in
      \eref{hashing} with $B=4$.  The variable nodes ($\bullet$) are the
      non-zero spectral values to be recovered. The white check nodes
      ($\square$) are the original time-domain samples. The colored squares are
      new check nodes after applying \algref{fasthadhash}.  
    }
    \flabel{bipartite}
  \end{figure*}

  \begin{figure*}
    \centering
    \includegraphics[width=0.8\linewidth]{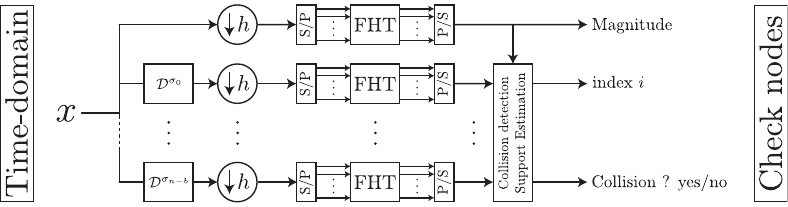}
    \caption{A block diagram of the SparseFHT algorithm in the time domain.
      The downsampling plus small size low complexity FHT blocks  compute
      different hash outputs. Delay blocks denote an index shift by $\sigma_i$
      before hashing.  The S/P and P/S  are serial-parallel and parallel-serial
      blocks to emphasize that the FHT operates on the whole signal at once.
      The collision detection/support estimation block implements
      \pref{collision_detect} to identify if there is a collision. Index $i$ is
      not valid when there is a collision.  
    }
    \label{hash_block}
  \end{figure*}

  \section{Sparse Fast Hadamard Transform}

  In this section, we give a brief overview of the main idea of Sparse Fast
  Hadamard Transform (SparseFHT). In particular, we explain the peeling decoder
  which recovers the non-zero spectral components and analyze its computational
  complexity.  \slabel{algoexact}

  \subsection{Explanation of the Algorithm}

  Assume that $x$ is an $N=2^n$ dimensional signal with a $K$-sparse WHT $X$,
  where $K=\bigO(N^\alpha)$ scales sub-linearly with $N$ with index $\alpha$.
  As $H_N^{-1}=H_N$, taking the inner product of the vector $X$ with the $i$-th
  row of the Hadamard matrix $H_N$ gives the time domain sample $x_i$.  Using
  the terminology of Coding theory, it is possible to consider the spectral
  components $X$ as variables nodes (information bits in coding theory) where
  the inner product of the $i$-th row of $H_N$ is like a parity check
  constraint on $X$. For example, the first row of the Hadamard matrix is the
  all-one vector which implies that the sum of all of the components of $X$
  must be equal to the first time domain sample. A similar interpretation holds
  for the other rows. Thus, the WHT can be imagined as a code over a bipartite
  graph. With this picture in mind, one can consider the recovery of the
  non-zero spectral values as a decoding problem over this bipartite graph. If
  we consider the WHT, it is easy to see that the induced bipartite graph on
  the non-zero spectral values is a complete (dense) bipartite graph because
  any variable node is connected to all of the check nodes as has been depicted
  in the left part of \ffref{bipartite}, where $\{X_1,X_8,X_{11}\}$ are the
  only non-zero variables in the spectral domain and each check constraint
  correspond to the value of a time domain sample. It is also implicitly assumed
  that one knows the support of $X$, $\{1,8,11\}$ in our case. At the moment,
  it is not clear how one can obtain the position of the non-zero variables. As
  we will explain, in the final version of the algorithm this can be done by
  using \pref{collision_detect}.

  For codes on bipartite graphs, there is a collection of low complexity belief
  propagation algorithms to recover the variable nodes given the value of check
  nodes. Most of these algorithms perform well assuming the  sparsity
  of the underlying bipartite graph. Unfortunately, the graph corresponding
  to WHT is dense, and probably not suitable for any of these belief propagation
  algorithms to succeed. 

  As explained in \sref{hashing}, by subsampling the time domain components of
  the signal it is possible to hash the spectral components in different bins
  as depicted for the same signal $X$ in the right part of  \ffref{bipartite}.
  The advantage of the hashing operation must be clear from this picture. The
  idea is that hashing \textit{`sparsifies'} the underlying bipartite graph.
  It is also important to notice that in the bipartite graph induced by
  hashing, one can obtain all of the values of parity checks (hash outputs) by
  using low complexity operations on a small set of time domain samples as
  explained in \pref{hashing}.

  We propose the following iterative algorithm to recover the non-zero spectral
  variables over the bipartite graph induced by hashing. The algorithm first
  tries to find a degree one check node. Using the terminology of
  \cite{Pawar:2013vm}, we call such a check node a \textit{singleton}. Using
  \pref{collision_detect}, the algorithm is able to find the position and the
  value of the corresponding non-zero component and, thus the algorithm can
  subtract (peel off) this variable from all other check nodes that are
  connected to it. In particular, after this operation the corresponding
  singleton check node gets value zero, namely, it is satisfied.  Equivalently,
  we can update the underlying graph by removing the mentioned variable node
  from the graph along with all the edges connected to it. This creates  an
  isolated (degree zero) check node which we call a \textit{zeroton}. Also
  notice that by removing some of the edges from the graph, the degree of the
  associated checks decreases by one, thus there is a chance that another
  singleton be found.

  The algorithm proceeds to peel off a singleton at a time until all of the
  check nodes are \textit{zeroton} (decoding succeeds) or all of the remaining
  check nodes have degree greater than one (we call them \textit{multiton}) and
  the algorithm fails to identify all of the non-zero spectral values.

  A more detailed pseudo-code of the proposed iterative algorithm is given in
  \algref{sfht}.

  \begin{algorithm}
    \caption{SparseFHT$(x, N, K, C, L, \fat{\Sigma})$}
    \alabel{sfht}
    \begin{algorithmic}
      \REQUIRE Input signal $x$ of length $N=2^n$. Sparsity $K$. Hash count $C$. Number of iterations of decoder $L$. 
      Array $\fat{\Sigma}$ of $C$ matrices in $\GL$, $\Sigma_c = [\sigma_{c,1}\,|\,\cdots\,|\,\sigma_{c,n}]$, $\sigma_{c,i}\in\F_2^n$.
      \ENSURE $X$ contains the sparse Hadamard spectrum of $x$.
      \STATE $B = \bigO(K)$
      \STATE $D = n-b+1$
      \FOR{$c=1,\ldots,C$}
      \STATE $U_{c,0} = \operatorname{FastHadamardHashing}(x, N, \Sigma_c, 0, B)$
      \FOR{$d=1,\ldots,D$}
      \STATE $U_{c,d} = \operatorname{FastHadamardHashing}(x, N, \Sigma_c, \sigma_{c,d}, B)$
      \ENDFOR
      \ENDFOR

      \FOR{$l=1,\ldots,L$}
      \FOR{$c=1,\ldots,C$}
      \FOR{$k=0,\ldots,B-1$}

      \IF{$U_{c,0,k} = 0$}
      \STATE continue to next $k$
      \ENDIF

      \STATE $\hat{v} \gets 0$

      \FOR{$d=1,\ldots,D$}
      \IF{$U_{c,d,k}/U_{c,0,k} = -1$}
      \STATE $\hat{v}_{d-1} \gets 1$
      \ELSIF{$U_{c,d,k}/U_{c,0,k} \neq 1$}
      \STATE continue to next $k$
      \ENDIF
      \ENDFOR

      \STATE $i \gets \Sigma_c^{-T}(\Psi_b\,k + \hat{v})$
      \STATE $X_i \gets U_{c,0,k}$

      \FOR{$c^\prime=1,\ldots,C$}
      \STATE $j \gets \Psi_b^T\Sigma_{c^\prime}^T\,i$
      \STATE $U_{c^\prime, 0, j} \gets U_{c^\prime, 0, j} - X_i$
      \FOR{$d^\prime=1,\ldots,D$}
      \STATE $U_{c^\prime, d^\prime, j} \gets U_{c^\prime, d^\prime, j} - X_i (-1)^{\ip{\sigma_{c^\prime,d^\prime}}{i}}$
      \ENDFOR
      \ENDFOR

      \ENDFOR
      \ENDFOR
      \ENDFOR
    \end{algorithmic}
  \end{algorithm}

  %
  %

  \subsection{Complexity Analysis}
  \slabel{complexity_analysis}

  Figure \ref{hash_block} shows a full block diagram of the SparseFHT
  algorithm. Using this block diagram, it is possible to prove part $1$ and $2$
  of Theorem \ref{main_theorem} about the sample and the computational
  complexity of the SparseFHT algorithm. The proof of the last part of Theorem
  \ref{main_theorem}, regarding the success probability of the algorithm, is
  the subject of Sections~\ref{sec:verysparse} and \ref{sec:lesssparse} for the
  very and less sparse regimes, respectively.

  {\bf Computational Complexity:}  As will be further clarified in
  Sections~\ref{sec:verysparse} and \ref{sec:lesssparse}, depending on the
  sparsity index of the signal $\alpha$, we will use $C$ different hash blocks,
  where $C \leq (\frac{1}{\alpha} \vee \frac{1}{1-\alpha})+1$ each with $B=2^b$
  different output bins. We always select $B=K$ to keep the average number of
  non-zero components per bin $\beta=\frac{K}{B}$ equal to $1$. This implies
  that computing the hash outputs via an FHT block of size $B$ needs $B
  \log_2(B)$ operations which assuming $K=B$, has a computational complexity $K
  \log_2(K)$. Moreover, we need to compute any hash output with
  $n-b=\log_2(\frac{N}{B})$ different shifts in order to do collision
  detection/support estimation, thus, the computational cost for each hash is
  $K \log_2(K) \log_2(\frac{N}{K})$. As we need to compute $C$ different hash
  blocks, the total computational complexity for each iteration will be $C K
  \log_2(K) \log_2(\frac{N}{K})$. We will explain later that the algorithm
  terminates in a fixed number of iterations independent of the value of
  $\alpha$ and the dimension of the signal $N$. Therefore, the total
  computational complexity of the algorithm will be $\bigO(C K \log_2(K)
  \log_2(\frac{N}{K}))$. 

  {\bf Sample Complexity:} Assuming $K=B$, computing each hash with $n-b$
  different shifts needs $K \log_2(\frac{N}{K})$ time domain samples.
  Therefore, the total sample complexity will be $C K \log_2(\frac{N}{K})$.

  \section{Performance Analysis of the very Sparse Regime}\label{sec:verysparse}

  In this section, we consider the very sparse regime, where $0<\alpha\leq
  \frac{1}{3}$. In this regime, we show that assuming a random support model
  for non-zero spectral components and a careful design of hash functions, it
  is possible to obtain a random bipartite graph with variable nodes
  corresponding to non-zero spectral components and with check nodes
  corresponding to outputs of hash functions. We explicitly prove that
  asymptotically this random graph behaves similarly to the ensemble of LDPC
  bipartite graph. Running the peeling decoder to recover the spectral
  components is also equivalent to the belief propagation (BP) decoding for a
  binary erasure channel (BEC). Fortunately, there is a rich literature in
  coding theory about asymptotic performance of the BP decoder. Specially, it
  is possible to show that the error (decoding failure) probability can be
  asymptotically characterized by a \textit{`Density Evolution'} (DE) equation
  allowing a perfect analysis of the peeling decoder. 

  We use the following steps to rigorously analyze the performance of the
  decoder in the very sparse regime:
  \begin{enumerate}
    \item We explain construction of suitable hash functions depending on the
      value of $\alpha \in (0, \frac{1}{3}]$.
    \item We rigorously analyze the structure of the induced bipartite graph
      obtained by treating the non-zero spectral components as variable
      nodes and the output of hash functions as check nodes. In particular,
      we prove that the resulting graph is a fully random left regular
      bipartite graph similar to the regular LDPC ensemble. We also obtain
      variable and check degrees distribution polynomials for this  graph.


    \item At every stage, the peeling decoder recovers some of the variable
      nodes, removing all the edges incident to those variable nodes. We use
      Wormald's method given in \cite{Wormald:1995wg} to prove the
      concentration of the number of unpeeled edges around its expected
      value, which we also characterize.  Wormald's method as exploited in
      \cite{Luby:2001iv}, uses the differential equation approach to track
      the evolution of the number of edges in the underlying bipartite
      graph. Specifically, it shows that the number of edges at every
      step of the algorithm is very well concentrated around the solution
      of the associated differential equations. 

    \item Wormald's method gives a concentration bound to the number of
      remaining edges as far as their count is a fixed ratio $\gamma \in
      (0,1)$ of the initial edges in the graph. Another expander argument
      as in \cite{Luby:2001iv} is necessary to show that if the peeling
      decoder peels $1-\gamma$ ratio of the edges successfully, it can
      continue to peel off all the remaining edges with very high
      probability.
  \end{enumerate}

  \subsection{Hash Construction}\label{hash_verysparse}

  For the very sparse regime, $0<\alpha\leq\frac{1}{3}$, consider those values
  of $\alpha$ equal to $\frac{1}{C}$ for some positive integer $C \geq 3$. We
  will explain later how to cover the intermediate values. For
  $\alpha=\frac{1}{C}$, we will consider $C$ different hash functions as
  follows. Let $x$ be an $N$ dimensional time domain signal with a WHT $X$,
  where $N=2^n$ and let $b=\frac{n}{C}$. As we explained before, the components
  of the vector $X$ can be labelled by $n$ dimensional binary vector from
  $\F_2^n$.  We design $C$ different subsampling operator, where the $i$-th one
  returns all of the variables with indices $i\, b$ up to $(i+1)b-1$ kept and
  the other indices set to zero. Using the terminology of \pref{hashing}, we
  let $\Sigma_i$ be the identity matrix with columns circularly shifted by
  $(i+1)b$ to the left.  Then, the hash operator given by \eref{hashing} is
  \begin{align*}
    \Hash_i = \Psi_b^T\Sigma_i^T = [ \mathbf{0}_{b \times i b}\ I_{b}\ \mathbf{0}_{b \times (n-(i+1)b)}],
  \end{align*}
  where $I_{b}$ is the identity matrix of order $b$ and $\Phi_b$ is defined in
  \eref{ds_mat}. To give further intuition about hash construction, let us
  label the elements of the vector $x$ with their binary representation
  $x_0^{n-1} \in \F_2^n$. Equivalent to the $C$ different subsampling
  operators, we can consider functions $h_i, i\in [C]$ where
  $h_i(x_0^{n-1})=(x_{i\, b}, x_{i\, b+1}, \dots, x_{ i\,b + b-1})$. The
  important point is that with this construction, the outputs of different
  $h_i$ depend on non overlapping portions of the labeling binary indices. In
  particular, labeling  the transform domain components by $X_0^{n-1} \in
  \F_2^n$ and ignoring the multiplicative constants, it is seen from Equation
  \eref{hash_bucket} that every spectral component $X_0^{n-1}$ is hashed to the
  bin labelled with $h_i(X_0^{n-1}) \in \F_2^b$ in hash $i$.

  In terms of complexity, to obtain the output of each hash bin, we only need
  to compute the WHT of a smaller subsampled signal of dimension $B$. Note that
  by hash construction, $K=B$ which implies that  all of the hash functions can
  be computed in $C K \log_2(K)$ operations. As we will explain later, we need
  at least $C=3$ hashes for the peeling algorithm to work successfully and that
  is the main reason why this construction works for $\alpha \leq \frac{1}{3}$.
  For intermediate values of $\alpha$, those not equal to $\frac{1}{C}$ for
  some integer $C$, one can construct $[\frac{1}{\alpha}]$ hashes with $B=2^{[n
    \alpha]}$ output bins and one hash with smaller number of output bins, thus
  obtaining a computational complexity of order $(1+[\frac{1}{\alpha}]) K
  \log_2(K)$.

  \subsection{Random Bipartite Graph Construction} 

  \subsubsection{Random Support Model}

  For an $N$ dimensional signal $x \in \R^N$, the support of $x$ is defined as
  $\supp(x)=\{i\in[N]: x_i \neq 0\}$. The signal $x$ is called $K$ sparse if
  $|\supp(x)|=K$, where for $A\subset [N]$, $|A|$ denotes the cardinality of $A$.
  For a given $(K,N)$, $\text{RS1}(K,N)$ is the class of all stochastic signals
  whose support is uniformly at random selected from the set of all ${N \choose
    K}$ possible supports of size $K$. We do not put any constraint on the
  non-zero components; they can be deterministic or random.  Model RS1 is
  equivalent to selecting $K$  out of $N$ objects at random without replacement.
  If we assume that the selection of the indices for the support is done
  independently but with replacement, we obtain another model that we call
  $\text{RS2}(K,N)$. In particular, if $V_i, i\in [K]$ are i.i.d. random
  variables uniformly distributed over $[N]$, a random support in
  $\text{RS2}(K,N)$ is given by the random set $\{V_i : i \in [K]\}$. Obviously,
  the size of a random support in $\text{RS2}(K,N)$ is not  necessarily fixed but
  it is at most $K$. The following proposition, proved in Appendix
  \ref{proof_randomsupp}, shows that in the sub-linear sparsity regime, RS1 and
  RS2 are essentially equivalent.  

  \begin{proposition}\label{randomsupp}
    Consider the random support model $\text{RS2}(K,N)$, where $K=N^\alpha$ for
    some fixed $0<\alpha <1$ and let $H$ be the random size of the support set.
    Asymptotically as $N$ tends to infinity $\frac{H}{K}$ converges to $1$ in
    probability.
  \end{proposition}

  \subsubsection{\textit{`Balls and Bins'} Model $\G(K,B,C)$} \label{BBM}

  Consider $C$ disjoint sets of check nodes $S_1,S_2,\dots,S_C$ of the same
  size $|S_i|=B$. A graph in the ensemble of  random bipartite graphs $\G$ with
  $K$ variable nodes at the left and $C \times B$ check nodes $\cup_{i=1}^C
  S_i$ at the right is generated as follows. Each variable node $v$ in $\G$,
  independently from other variable nodes, is connected to check nodes
  $\{s_1,s_2,\dots, s_C\}$ where $s_i \in S_i$ is uniformly at random selected
  from $S_i$ and selection of $s_i$'s are independent of one another. Every
  edge $e$ in $\G$ can be labelled as $(v,c)$, where $v \in [K]$ is a variable
  node and $c$ is a check node in one of $S_1,S_2,\dots,S_C$. For a variable
  node $v$, the neighbors of $v$ denoted by ${\cal N}(v)$ consists of $C$
  different check nodes connected to $v$, each of them from a different $S_i$.
  Similarly, for a check node $c \in \cup_{i=1}^C S_i$, ${\cal N}(c)$ is the
  set of all variable nodes connected to $c$.  

  By construction, all of the  resulting bipartite graphs in the ensemble are
  left regular with variable degree $C$ but the check node degree is not fixed.
  During the construction, it might happen that two variable nodes have exactly
  the same neighborhood. In that case, we consider them as equivalent variables
  and keep only one of them and remove the other, thus the number of variable
  nodes in a graph from the ensemble $\G(K,B,C)$ might be less than $K$. 

  This model is a variation of the Balls and Bins model, where we have $K$
  balls, $C$ buckets of different color each containing $B$ bins and every ball
  selects one bin from each bucket at random independent of the other balls.

  Here we also recall some terminology from graph theory that we will use
  later. A walk of size $\ell$ in graph $\G$ starting from a node $v\in [K]$ is
  a set of $\ell$ edges $e_1,e_2, \dots,e_\ell$, where $v \in e_1$ and where
  consecutive edges are different, $e_i \neq e_{i+1}$, but incident with each
  other $e_i \cap e_{i+1} \neq \emptyset$. A directed neighborhood of an edge
  $e=(v,c)$ of depth $\ell$ is the induced subgraph in $\G$ consisting of all
  edges and associated check and variable nodes in all walks of size $\ell+1$
  starting from $v$ with the first edge being $e_1=(v,c)$. An edge $e$ is said
  to have a tree neighborhood of depth $\ell$ if the directed neighborhood of
  $e$ of depth $\ell$ is a tree.

  \subsubsection{Ensemble of Graphs Generated by Hashing}

  In the very sparse regime ($0<\alpha<\frac{1}{3}$), in order to keep the
  computational complexity of the hashing algorithm around $O(K \log_2(K))$, we
  constructed $C=\frac{1}{\alpha}$ different surjective hash functions $h_i:
  \F_2^n \to \F_2^b$, $i \in [C]$, where $b\approx n \alpha$ and where for an
  $x \in \F_2^n$ with binary representation $x_0^{n-1}$,
  $h_i(x_0^{n-1})=(x_{i\, b}, x_{i\, b+1}, \dots, x_{ i\,b + b-1})$. We also
  explained that in the spectral domain, this operation is equivalent to
  hashing spectral the component labeled with $X_0^{n-1} \in \F_2^n$ into the
  bin labelled with $h_i(X_0^{n-1})$. Notice that by this hashing scheme there
  is a one-to-one relation between a spectral element $X$  and its bin indices
  in different hashes $(h_0(X),h_1(X),\dots,h_{C-1}(X))$. 

  Let $V$ be a uniformly distributed random variable over $\F_2^n$. It is easy
  to check that in the binary representation of $V$, $V_0^{n-1}$ are like
  i.i.d. unbiased bits. This implies that $h_0(V), h_1(V), \dots, h_{C-1}(V)$
  will be independent from one another because they depend on disjoint subsets
  of $V_0^{n-1}$. Moreover, $h_i(V)$ is also uniformly distributed over
  $\F_2^b$. 

  Assume that $X_1, X_2, \dots, X_K$ are $K$ different variables in $\F_2^n$
  denoting the position of non-zero spectral components. For these $K$
  variables and hash functions $h_i$, we can associate a bipartite graph as
  follows. We consider $K$ variable nodes corresponding to $X_1^K$ and $C$
  different set of check nodes $S_0,S_1, \dots , S_{C-1}$ each of size $B=2^b$.
  The check nodes in each $S_i$ are labelled by elements of $\F_2^b$. For each
  variable $X_i$ we consider $C$ different edges connecting $X_i$ to check
  nodes labelled with $h_j(X_i) \in S_j$, $j\in [C]$. 

  \begin{proposition}
    \plabel{BBequiv}
    Let $h_i: \F_2^n \to \F_2^b$, $i\in[C]$ be as explained before. Let $V_1,
    V_2, \dots , V_K$ be a set of variables generated from the ensemble
    $\text{RS2}(K,N)$, $N=2^n$ denoting the position of non-zero components.
    The bipartite graph associated with variables $V_1^K$ and hash functions
    $h_i$ is a graph from ensemble $\G(K,B,C)$, where $B=2^b$. 
  \end{proposition}

  \begin{IEEEproof}
    As $V_1^K$ belong to the ensemble $\text{RS2}(N,K)$, they are i.i.d.
    variables uniformly distributed in $[N]$. This implies that for a specific
    $V_i$, $h_j(V_i)$, $j \in [C]$ are independent from one another. Thus,
    every variable node selects its neighbor checks in $S_0,S_1, \dots,
    S_{C-1}$ completely at random. Moreover, for any $j \in [C]$, the variables
    $h_j(V_1), \dots, h_j(V_K)$ are also independent, thus each variable
    selects its neighbor checks in $S_j$ independent of all other variables.
    This implies that in the corresponding bipartite graph, every variable node
    selects its $C$ check neighbors completely at random independent of other
    variable nodes, thus it belongs to  $\G(K,B,C)$.
  \end{IEEEproof}

  In \sref{algoexact}, we explained the peeling decoder over the bipartite
  graph induced by the non-zero spectral components. It is easy to see that the
  performance of the algorithm always improves if we remove some of the
  variable nodes from the graph because it potentially reduces the number of
  colliding variables in the graph and there is more chance for the peeling
  decoder to succeed decoding. 

  \begin{proposition}\label{graph_upperbound}
    Let $\alpha$, $C$, $K$, $h_i, i\in[C]$ be as explained before. Let $\G$ be
    the bipartite graph induced by the random support set $V_1^K$ generated
    from RS1 and hash functions $h_i$. For any $\epsilon>0$, asymptotically as
    $N$ tends to infinity, the average failure probability of the peeling
    decoder over $\G$ is upper bounded by its average failure probability  over
    the ensemble $\G(K(1+\epsilon),B,C)$. 
  \end{proposition} 

  \begin{IEEEproof}
    Let $\G_\epsilon$ be a graph from ensemble $\G(K(1+\epsilon),B,C)$. From
    Proposition \ref{randomsupp}, asymptotically the number of variable nodes
    in $\G_\epsilon$ is greater than $K$. If we drop some of the
    variable nodes at random from $\G_\epsilon$ to keep only $K$ of them we obtain a
    graph from ensemble $\G$. From the explanation of the peeling decoder, it
    is easy to see that the performance of the decoder improves by removing
    some of the variable nodes because in that case less variables are collided
    together in different bins and there is more chance to peel them off. This
    implies that peeling decoder performs strictly better over $\G$ compared
    with $\G_\epsilon$. 
  \end{IEEEproof}

  \begin{remark}
    If we consider the graph induced by $V_1^K$ from RS1 and hash functions
    $h_i$, the edge connection between variable nodes and check nodes is not
    completely random thus it is not compatible with Balls-and-Bins model
    explained before. Proposition \ref{graph_upperbound} implies that
    asymptotically the failure probability for this model can be upper bounded
    by the failure probability of  the peeling decoder for Balls-and-Bins model
    of slightly higher number of edges $K(1+\epsilon)$.
  \end{remark}

  \subsubsection{Edge Degree Distribution Polynomial}

  As we explained in the previous section, assuming a random support model for
  non-zero spectral components in the very sparse regime
  $0<\alpha<\frac{1}{3}$, we obtained a random graph from ensemble $\G(K,B,C)$.
  We also assumed that $n \alpha \in \mN$ and we selected $b=n \alpha$, thus
  $K=B$. Let us call $\beta = \frac{K}{B}$ to be the average number of non-zero
  components per a hash bin. In our case, we designed hashes so that $\beta=1$.
  As the resulting bipartite graph is left regular, all of the variable nodes
  have degree $C$ whereas for a specific check node the degree is random and
  depends on the graph realization.

  \begin{proposition}\label{check_poisson}
    Let $\G(K,B,C)$ be the random graph ensemble as before with
    $\beta=\frac{K}{B}$ fixed. Then asymptotically as  $N$ tends to infinity
    the check degree converges to a Poisson random variable with parameter
    $\beta$.
  \end{proposition}

  \begin{IEEEproof}
    Construction of the ensemble $\G$ shows that any variable node has a
    probability of $\frac{1}{B}$ to be connected to a specific check node, $c$,
    independent of all other variable nodes. Let $Z_i \in \{0,1\}$ be a
    Bernoulli random variable where $Z_i=1$ if and only if variable $i$ is
    connected to check node $c$. It is easy to check that the degree of $c$
    will be $Z=\sum_{i=1}^K Z_i$. The Characteristic function of $Z$ can be
    easily obtained:
    \begin{align*}
      \Phi_Z(\omega)&=\E{e^{j \omega Z}}=\prod _{i=1}^K \E{e^{j \omega Z_i}}\\
      &=\left(1+ \frac{1}{B}(e^{j \omega} -1)\right)^{\beta B} \to e^{\beta (e^{j \omega}-1)},
    \end{align*}
    showing the convergence of $Z$ to a Poisson distribution with parameter $\beta$.
  \end{IEEEproof}

  For a bipartite graph, the edge degree distribution polynomial is defined by
  $\rho(\alpha)=\sum_{i=1}^\infty \rho_i \alpha^{i-1}$ and
  $\lambda(\alpha)=\sum_{i=1} ^ \infty \lambda_i \alpha^{i-1}$, where $\rho_i$
  ($ \lambda_i$) is the ratio of all edges that are connected to a check node
  (variable node) of degree $i$. Notice that we have $i-1$ instead of $i$ in
  the formula. This choice makes the analysis to be written in a more compact
  form as we will see. 

  \begin{proposition}
    Let $\G$ be a random bipartite graph from the ensemble $\G(K,B,C)$ with
    $\beta=\frac{K}{B}$. Then $\lambda(\alpha)=\alpha^{C-1}$ and $\rho(\alpha)$
    converges to $e^{-\beta(1-\alpha)}$ as $N$ tends to infinity.
  \end{proposition} 

  \begin{IEEEproof}
    From left regularity of a graph from ensemble $\G$, it results that all of
    the edges are connected to variable nodes of degree $C$, thus
    $\lambda(\alpha)=\alpha^{C-1}$ and the number of edges is equal to $C \,K$.
    By symmetry of hash construction, it is sufficient to obtain the edge
    degree distribution polynomial for check nodes of the first  hash. The
    total  number of edges that are connected to the check nodes of the first
    hash is equal to $K$. Let $N_i$ be the number of check nodes in this hash
    with degree $i$. By definition of $\rho_i$, it results that 
    \begin{align*}
      \rho_i=\frac{i N_i}{K}= \frac{i\,N_i/B}{K/B}.
    \end{align*}
    Let $Z$ be the random variable as in the proof of Proposition
    \ref{check_poisson} denoting the degree of a specific check node. Then, as
    $N$ tends to infinity one can show that 
    \begin{align*}
      \lim_{N \to \infty} \frac{N_i}{B}=\lim_{N \to \infty} \prob{Z=i}=\frac{e^{-\beta} \beta^i}{i!}\ \text{a.s.}
    \end{align*}
    Thus $\rho_i$ converges almost surely to $\frac{e^{-\beta}
      \beta^{i-1}}{(i-1)!}$. As $\rho_i \leq 1$, for any $\alpha:
    |\alpha|<1-\epsilon$, $|\rho_i \alpha^{i-1}| \leq (1-\epsilon)^{i-1}$ and
    applying the Dominated Convergence Theorem, $\rho(\alpha)$ converges to
    $\sum_{i=1}^\infty \frac{e^{-\beta} \beta^{i-1}}{(i-1)!} \alpha^{i-1} =
    e^{-\beta(1-\alpha)}$.
  \end{IEEEproof}

  \subsubsection{Average Check Degree Parameter $\beta$}

  In the very sparse regime, as we explained assuming that $b=n \alpha$ is an
  integer we designed independent hashes with $B=2^b$ output bins so that
  $\beta=\frac{K}{B}=1$. As we will see the performance of the peeling decoder
  (described later by the DE equation in \eqref{prob_iteration}) depends on the parameter $\beta$. The less $\beta$
  the better the performance of the peeling decoder. Also notice that
  decreasing $\beta$ via increasing $B$ increases the time complexity $O(B
  \log_2(B))$ of computing  the hash functions. For the general case, one can
  select $B$ such that $\beta \in [1,2)$ or at the cost of increasing the
  computational complexity make $\beta$ smaller for example $\beta\in
  [\frac{1}{2},1)$ to obtain a better performance. 


  \subsection{Performance Analysis of the Peeling Decoder}

  Assume that $\G$ is the random bipartite graph resulting from applying $C$
  hashes to signal spectrum. As we explained in \sref{algoexact}, the iterative
  peeling algorithm starts by finding a singleton (check node of degree $1$
  which contains only one variable node or non-zero spectral components). The
  decoder peels off this variable node and removes all of the edges connected
  to it from the graph. The algorithm continues by peeling off a singleton at
  each step until all of the check nodes are zeroton; all of the non-zero
  variable nodes are decoded, or all of the remaining unpeeled check nodes are
  multiton in which case the algorithm fails to completely decode all the
  spectral variables. 

  \subsubsection{{\bf Wormald's Method}}

  In order to analyze the behavior of the resulting random graphs under the
  peeling decoding, the authors in \cite{Luby:2001iv} applied Wormald's method
  to track the ratio of edges in the graph connected to check nodes of degree
  $1$ (singleton). The essence of Wormald's method is to approximate the
  behavior of a stochastic system (here the random bipartite graph), after
  applying suitable time normalization,  by a deterministic differential
  equation. The idea is that asymptotically as the size of the system becomes
  large (thermodynamic limit), the random state of the system is, uniformly for
  all times during the run of the algorithm, well concentrated around the
  solution of the differential equation. In \cite{Luby:2001iv}, this method was
  applied to analyze the performance of the peeling decoder for bipartite graph
  codes over the BEC.  We briefly explain the problem setting in
  \cite{Luby:2001iv} and how it can be used in our case.

  Assume that we have a bipartite graph $\G$ with $k$ variable nodes at the
  left, $c\, k$ check nodes at the right and with edge degree polynomials
  $\lambda(x)$ and $\rho(x)$. We can define a channel code ${\cal C}(\G)$ over
  this graph as follows. We assign $k$ independent message bits to  $k$ input
  variable nodes. The output of each check node is the module $2$ summation
  (XOR or summation over $\F_2$) of the all of the message bits that are
  connected to it. Thus, the resulting code will be a systematic code with $k$
  message bits along with $c\,k$ parity check bits. To communicate a $k$ bit
  message over the channel, we send $k$ message bits and all of the check bits
  associated with them. While passing through the BEC, some of the message bits
  or check bits are erased independently. Assume a specific case in which the
  message bits and check bits are erased independently with probability
  $\delta$ and $\delta '$ respectively. Those message bits that pass perfectly
  through the channel are successfully transmitted, thus, the decoder tries to
  recover the erased message bits from the redundant information received via
  check bits. If we consider the induced graph after removing all variable
  nodes and check nodes corresponding to the erased ones from $\G$, we end up
  with another bipartite graph $\G'$. It is easy to see that over the new graph
  $\G'$, one can apply the peeling decoder to recover the erased bits.

  In \cite{Luby:2001iv}, this problem was fully analyzed for the case of
  $\delta'=0$, where all of the check bits are received perfectly but $\delta$
  ratio of the message bits are erased independently from one another. In other
  words, the final graph $\G'$ has on average $k \delta$ variable nodes to be
  decoded. Therefore, the analysis can be simply applied to our case, by
  assuming that $\delta \to 1$, where all of the variable nodes are erased
  (they are all unknown and need to by identified). Notice that from the
  assumption $\delta'=0$ no check bit is erased as is the case in our problem.
  In particular, Proposition $2$ in \cite{Luby:2001iv} states that

  {\bf Proposition 2 in \cite{Luby:2001iv}:} Let $\G$ be a bipartite graph
  with edge degrees specified by $\lambda(x)$ and $\rho(x)$ and with $k$
  message bits chosen at random. Let $\delta$ be fixed so that
  \begin{align*}
    \rho(1-\delta \lambda(x)) > 1- x, \ \ \text{for } x \in (0,1].
  \end{align*}
  For any $\eta>0$, there is some $k_0$ such that for all $k>k_0$, if the
  message bits of ${\cal C}(\G)$ are erased independently with probability
  $\delta$, then with probability at least $1-k^{\frac{2}{3}}
  \exp(-\sqrt[3]{k}/2)$ the recovery algorithm terminates with at most $\eta k$
  message bits erased.

  Replacing $\delta=1$ in the proposition above, we obtain the following
  performance guarantee for the peeling decoder. 

  \begin{proposition}\label{peel_performance}
    Let $\G$ be a bipartite graph from the ensemble $\G(K,B,C)$ induced by
    hashing functions $h_i, i \in[C]$ as explained before with
    $\beta=\frac{K}{B}$ and edge degree polynomials $\lambda(x)=x^{C-1}$ and
    $\rho(x)=e^{-\beta(1-x)}$ such that 
    \begin{align*}
      \rho(1- \lambda(x)) > 1- x, \ \ \text{for } x \in (0,1].
    \end{align*}
    Given any $\epsilon \in (0,1)$, there is a $K_0$ such that for any $K>K_0$
    with probability at least $1-K^{\frac{2}{3}} \exp(-\sqrt[3]{K}/2)$ the
    peeling decoder terminates with at most $\epsilon \, K$ unrecovered
    non-zero spectral components.
  \end{proposition}

  Proposition \ref{peel_performance} does not guarantee the success of the
  peeling decoder. It only implies that with very high probability, it can peel
  off any ratio $\eta \in(0,1)$ of non-zero components but not necessarily all
  of them. However, using a combinatorial argument, it is possible to prove
  that with very high probability any graph in the ensemble $\G$ is an
  expander graph, namely, every small enough subset of left nodes has many check
  neighbors. This implies that if the peeling decoder can decode a specific
  ratio of variable nodes, it can proceed to decode all of them. A slight
  modification of Lemma $1$ in \cite{Luby:2001iv} gives the following result
  proved in Appendix \ref{expander_arg}.

  \begin{proposition}\label{expander}
    Let $\G$ be a graph from the ensemble $\G(K,B,C)$ with $C\geq 3$. There is
    some $\eta>0$ such that with probability at least
    $1-O(\frac{1}{K^{3(C/2-1)}})$, the recovery process restricted to the
    subgraph induced by any $\eta$-fraction of the left nodes terminates
    successfully. 
  \end{proposition}

  {\bf Proof of Part 3 of Theorem \ref{main_theorem} for $\alpha \in (0,\frac{1}{3}]$:} \vspace{0.5mm}
  In the very sparse regime $\alpha \in (0,\frac{1}{3}]$, we construct
  $C=[\frac{1}{\alpha}] \geq 3$ hashes each containing $2^{n \alpha}$ output
  bins. Combining Proposition \ref{peel_performance} and \ref{expander}, we
  obtain that the success probability of the peeling decoder is lower bounded
  by $1-O(\frac{1}{K^{3(C/2-1)}})$ as mentioned in Remark
  \ref{remark_verysparse}.

  \vspace{1mm}
  \subsubsection{{\bf Analysis based on Belief Propagation over Sparse Graphs}}

\begin{figure}[h]
    \centering
    \includegraphics[width=0.6\linewidth]{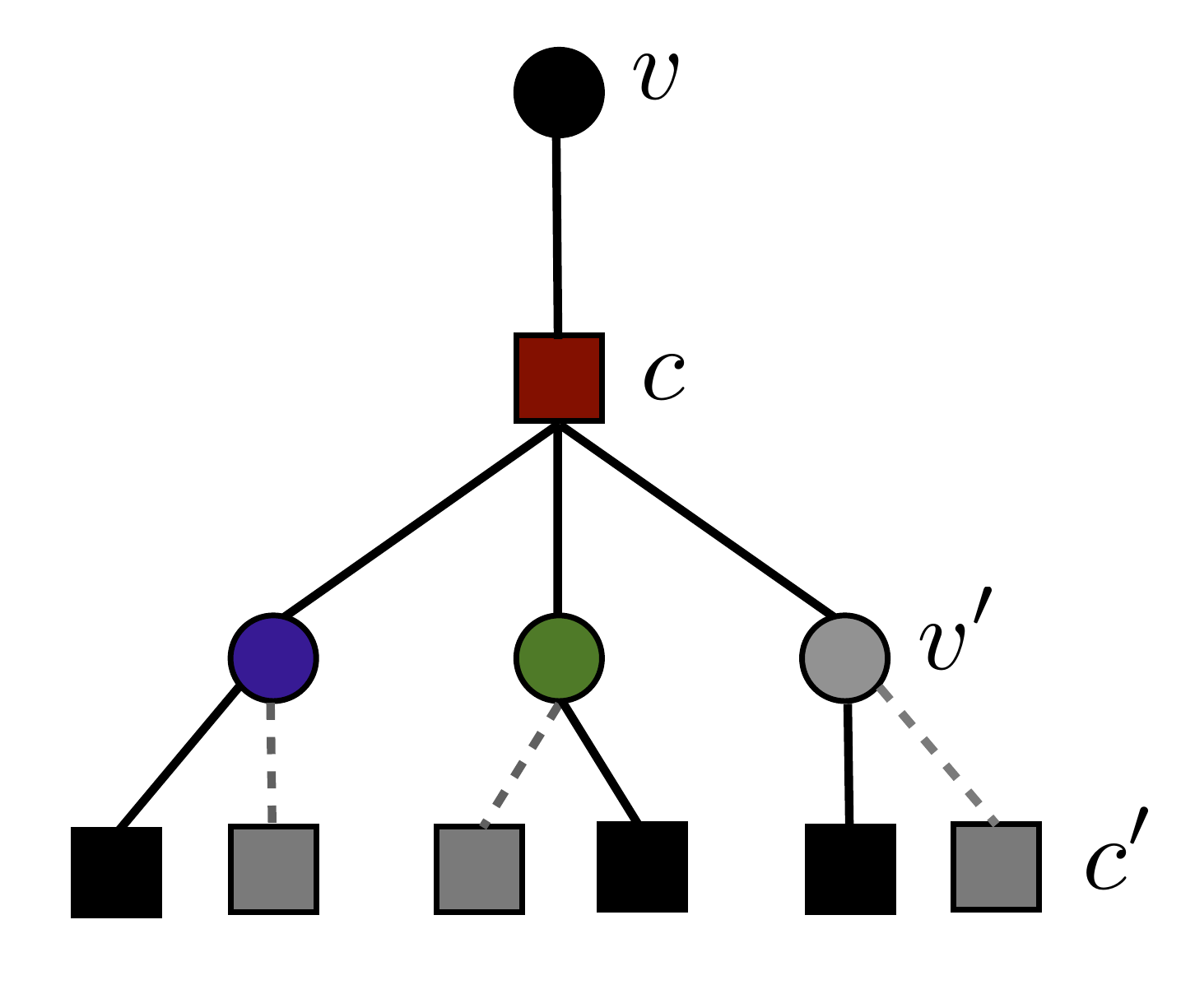}
    \caption{Tree-like neighborhood an an edge $e=(v,c)$. Dashed lines show the
      edges that have been removed before iteration $t$. The edge $e$ is peeled
      off at iteration $t$ because all the variable nodes $v'$ connected to $c$
      are already decoded, thus $c$ is a singleton check.
    }
    \label{tree_like}
  \end{figure}

  In this section, we give another method of analysis and further intuition
  about the performance of the peeling decoder and why it works very well in
  the very sparse regime. This method is based on the analysis of BP decoder
  over sparse locally tree-like graphs. The analysis is very similar to the
  analysis of the peeling decoder to recover non-zero frequency components in
  \cite{Pawar:2013vm}. Consider a specific edge $e=(v,c)$ in a graph from
  ensemble $\G(K,B,C)$. Consider a directed neighborhood of this edge of depth
  $\ell$ as explained is \ref{BBM}. At the first stage, it is easy to see that
  this edge is peeled off from the graph assuming that all of the edges
  $(c,v')$ connected to the check node $c$ are peeled off because in that case
  check $c$ will be a singleton allowing to decode the variable $v$. This
  pictorially shown in Figure \ref{tree_like}.

  One can proceed in this way in the directed neighborhood to find the
  condition under which the variable $v'$ connected to $c$ can be peeled off
  and so on. Assuming that the directed neighborhood is a tree, all of the
  messages that are passed from the leaves up to the head edge $e$ are
  independent from one another. Let $p_\ell$ be the probability that edge $e$
  is peeled off depending on the information received from the directed
  neighborhood of depth $\ell$ assuming a tree up to depth $\ell$.  A simple
  analysis similar to \cite{Pawar:2013vm}, gives the following recursion
  \begin{align}\label{prob_iteration}
    p_{j+1}=\lambda(1-\rho(1-p_j)), \ \ j \in [\ell],
  \end{align}
  where $\lambda$ and $\rho$ are the edge degree polynomials of the ensemble
  $\G$. This iteration shows the progress of the peeling decoder in recovering
  unknown variable nodes. In \cite{Pawar:2013vm}, it was proved that for any
  specific edge $e$, asymptotically with very high probability the directed
  neighborhood of $e$ up to any fixed depth $\ell$  is a tree. Specifically, if
  we start from a left regular graph $\G$ from $\G(K,B,C)$ with $K C$ edges,
  after $\ell$ steps of decoding, the average number of unpeeled edges is
  concentrated around $K C p_\ell$. Moreover, a martingale argument was applied
  in \cite{Pawar:2013vm} to show that not only the average of unpeeled edges is
  approximately $K C p_\ell$ but also with very high probability the number of
  those edges is well concentrated around $K C p_\ell$. 

\begin{figure}[h]
    \centering
    \includegraphics[width=0.8\linewidth]{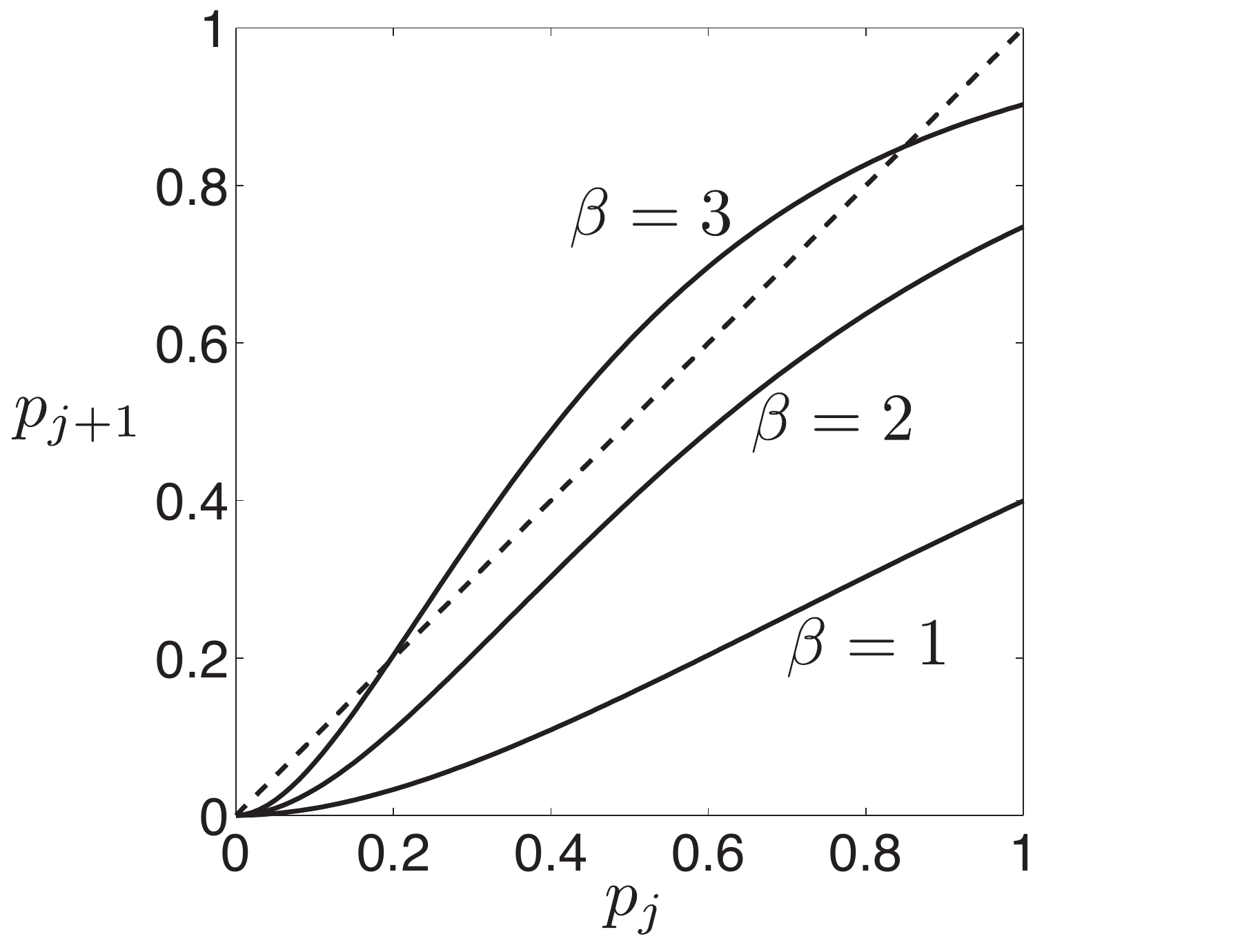}
    \caption{Density Evolution equation for $C=3$ and different values of $\beta=\frac{K}{B}$}
    \label{fixed_point}
  \end{figure}

  Equation \eqref{prob_iteration} is in general known as density evolution
  equation. Starting from $p_0=1$, this equation fully predicts the behavior of
  the peeling decoding over the ensemble $\G$. Figure~\ref{fixed_point} shows a
  typical behavior of this iterative equation for different values of the
  parameter $\beta=\frac{K}{B}$. 

  For very small values of $\beta$, this equation has only a fixed point $0$
  which implies that asymptotically the peeling decoder can recover a ratio of
  variables very close to $1$. However, for large values of $\beta$,
  i.e. $\beta \gtrsim 2.44$ for $C=3$, this equation has a fixed point greater
  than $0$. The largest fixed point is the place where the peeling decoder
  stops  and can not proceed to decode the remaining variables. It is easy to
  see that the only fixed point is $0$ provided that for any $p \in (0,1]$, $p
  > \lambda(1-\rho(1-p))$. As $\lambda :[0,1] \to [0,1]$, $\lambda(x)=x^{C-1}$
  is an increasing function of $x$, by change of variable $x=\lambda^{-1}(p)$,
  one obtains that $x>1-\rho(1-\lambda(x))$ or equivalently $$\rho(1-
  \lambda(x)) > 1- x, \ \ \text{for } x \in (0,1].$$
  This is exactly the same result that we obtained by applying Wormald's
  method as in \cite{Luby:2001iv}. In particular, this analysis clarifies the
  role of $x$ in Wormald's method. 

  Similar to Wormald's method, this analysis only guaranties that for any
  $\epsilon \in (0,1)$, asymptotically as $N$ tends to infinity, $1-\epsilon$
  ratio of the variable nodes can be recovered. An expander argument is again
  necessary to guarantee the full recovery of all the remaining variables. 

  \section{Performance Analysis of the Less Sparse Regime}\label{sec:lesssparse}

  For the less sparse regime ($\frac{1}{3} < \alpha <1$), similar to the very
  sparse case, we will first construct suitable hash functions which guarantee
  a low computational complexity of order $O(K \log_2(K) \log_2(\frac{N}{K}))$
  for the recovery of non-zero spectral values. Assuming a uniformly random
  support model in the spectral domain, similar to the very sparse case, we can
  represent the hashes by a regular bipartite graph. Over this graph, the
  peeling algorithm proceeds to find singleton checks and peel the associated
  variables from the graph until no singleton remains. The recovery is
  successful if all of the variables are peeled off, thus, all of the remaining
  checks are zeroton otherwise some of the non-zero spectral values are not
  recovered and the perfect recovery fails.

  As we will explain, the structure of the induced bipartite graph in this
  regime is a bit different than the very sparse one. The following steps are
  used to analyze the performance of the peeling decoder:
  \begin{enumerate}
    \item Constructing suitable hash functions
    \item Representing hashing of non-zero spectral values by an equivalent bipartite graph
    \item Analyzing the performance of the peeling decoder over the resulting bipartite graph
  \end{enumerate}
  For simplicity, we consider the case where $\alpha =1-\frac{1}{C}$ for some
  integer $C \geq 3$. We will explain how to deal with arbitrary values of $C$ and
  $\alpha$, especially those in the range $(\frac{1}{3},\frac{2}{3})$, in \sref{generalized_hash}. 

  \subsection{Hash Construction}

  Assume that $\alpha=1-\frac{1}{C}$ for some integer $C \geq 3$. Let $x$ be an
  $N$ dimensional signal with $N=2^n$ and let $X$ denote its WHT. For
  simplicity, we label the components of $X$ by a binary vector $X_0^{n-1} \in
  \F_2^n$. Let $t=\frac{n}{C}$ and let us divide the set of $n$ binary indices
  $X_0^{n-1}$ into $C$ non-intersecting subsets $r_0, r_1, \dots , r_{C-1}$,
  where $r_i=X_{i\, t} ^{(i+1)t -1}$. It is clear that there is a one-to-one
  relation between each binary vector $X_0^{n-1}\in \F_2^n$ and its
  representation $(r_0,r_1,\dots,r_{C-1})$. We construct $C$ different hash
  function $h_i, i \in[C]$ by selecting different subsets of
  $(r_0,r_1,\dots,r_{C-1})$ of size $C-1$ and appending them together. For
  example 
  \begin{align*}
    h_1(X_0^{n-1})=(r_0,r_1,\dots,r_{C-2})=X_0^{(C-1) t-1},
  \end{align*}
  and the hash output is obtained by appending $C-1$ first $r_i, i \in [C]$.
  One can simply check that $h_i, i\in[C]$ are linear surjective functions from
  $\F_2^n$ to $\F_2^b$, where $b=(C-1)t$. In particular, the range of each hash
  consists of $B=2^b$ different elements of $\F_2^b$. Moreover, if we denote
  the null space of $h_i$ by $\Null(h_i)$, it is easy to show that for any $i,j
  \in [C], i\neq j$, $\Null(h_i) \cap \Null(h_j)= \mathbf{0} \in \F_2^n$.

  Using the subsampling property of the WHT and similar to the hash
  construction that we had in Subsection \ref{hash_verysparse}, it is seen that
  subsampling the time domain signal and taking WHT of the subsampled signal is
  equivalent to hashing the spectral components of the signal. In particular,
  all of the spectral components $X_0^{n-1}$ with the same $h_i(X_0^{n-1})$ are
  mapped into the same bin in hash $i$, thus, different bins of the hash can be
  labelled with $B$ different elements of $\F_2^b$. 

  It is easy to see that, with this construction the average number of non-zero
  elements per bin in every hash is kept at $\beta=\frac{K}{B}=1$ and  the
  complexity of computing all the hashes along with their $n-b$ shifts, which
  are necessary for collision detection/support estimation, is $C K \log_2(K)
  \log_2(\frac{N}{K})$. The sample complexity can also be easily checked to be
  $C K \log_2(\frac{N}{K})$.

  \subsection{Bipartite Graph Representation}

\begin{figure}[h]
    \centering
    \includegraphics[width=0.95\linewidth]{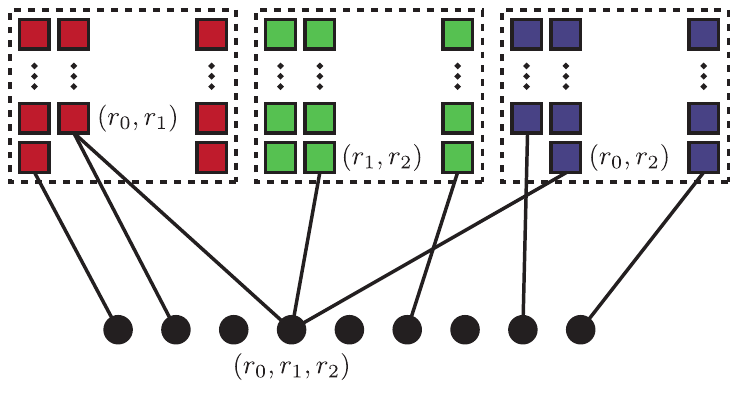}
    \caption{Bipartite graph representation for the less sparse case $\alpha=\frac{2}{3}$, $C=3$}
    \label{less_sparse_graph}
  \end{figure}

  Similar to the very sparse regime, we can assign a bipartite graph with the
  $K$ left nodes  (variable nodes) corresponding to non-zero spectral
  components and with $C B$ right nodes corresponding to different bins of all
  the hashes. In particular, we consider $C$ different set of check nodes
  $S_1,S_2, \dots,S_C$ each containing $B$ nodes labelled with the elements of
  $\F_2^b$ and a specific non-zero spectral component labelled with
  $X_0^{n-1}$ is connected to nodes $s_i \in S_i$ if and only if the binary
  label assigned to $s_i$ is $h_i(X_0^{n-1})$. In the very sparse regime, we
  showed that if the support of the signal is generated according to the
  $\text{RS2}(K,N)$, where $K$ random positions are selected uniformly at random
  independent from one another from $[N]$, then the resulting graph is a random
  left regular bipartite graph, where each variable nodes select its $C$
  neighbors in  $S_1,S_2, \dots,S_C$ completely independently. However, in the
  less sparse regime, the selection of the neighbor checks in different hashes
  is not completely random. To explain more, let us assume that
  $\alpha=\frac{2}{3}$, thus $C=3$. Also assume that for a non-zero spectral
  variable labelled with $X_0^{n-1}$, $r_i$ denotes $X_{i\, t}^{(i+1)t-1}$,
  where $t=\frac{n}{C}$. In this case, this variable is connected to bins
  labelled with $(r_0,r_1)$, $(r_1,r_2)$ and $(r_0,r_2)$ in $3$ different
  hashes. This has been pictorially shown in Figure \ref{less_sparse_graph}.

  If we assume that $X_0^{n-1}$ is selected uniformly at random from $\F_2^n$
  then the bin numbers is each hash, i.e. $(r_0,r_1)$ in the first hash, are
  individually selected uniformly at random among all possible bins. However, it
  is easily seen that the joint selection of bins is not completely random
  among different hashes. In other words, the associated bins in different
  hashes are not independent from one another. However, assuming the random
  support model, where $K$ variable $V_1^K$ are selected independently as the
  position of non-zero spectral variables, the bin association for different
  variables $V_i$ is still done independently. 

  \subsection{Performance Analysis of the Peeling Decoder}

  As the resulting bipartite graph is not a completely random graph, it is not
  possible to directly apply Wormald's method as we did for the very sparse
  case as in \cite{Luby:2001iv}. However, an analysis based on the DE for the
  BP algorithm can still be applied. In other words, setting $p_0=1$ and 
  \begin{align*}
    p_{j+1}=\lambda(1-\rho(1-p_j)), \ \ j \in [\ell],
  \end{align*} 
  as in \eqref{prob_iteration} with $\lambda$ and $\rho$ being the edge degree
  polynomials of the underlying bipartite graph, it is still possible to show
  that after $\ell$ steps of decoding the average number of unpeeled edges is
  approximately $K C p_\ell$. A martingale argument similar to
  \cite{Pawar:2013vm} can be applied to show that the number of remaining edges
  is also well concentrated around its average. Similar to the very sparse
  case, this argument asymptotically guarantees the recovery of any ratio of
  the variables between $0$ and $1$. Another argument is necessary to show that
  if the peeling decoder decodes a majority of the variables, it can proceed to
  decode all of them with very high probability. To formulate this, we use the
  concept of trapping sets for the peeling decoder.

  \begin{definition}
    Let $\alpha=1-\frac{1}{C}$ for some integer $C\geq 3$ and let $h_i,
    i\in[C]$ be a set of hash functions as explained before. A subset of
    variables $T \subset \F_2^n$ is called a trapping set for the peeling
    decoder if for any $v \in T$ and for any $i \in [C]$, there is another $v_i
    \in T$, $v\neq v_i$ such that $h_i(v)=h_i(v_i)$, thus colliding with $v$ in
    the $i$-th hash.
  \end{definition}

  Notice that a trapping set can not be decoded because all of its neighbor
  check nodes are multiton. We first analyze the structure of the trapping set
  and find the probability that a specific set of variables build a trapping
  set. Let $X$ be a spectral variable in the trapping set with the
  corresponding binary representation $X_0^{n-1}$ and assume that $C=3$. As we
  explained, we can equivalently represent this variable with $(r_0,r_1,r_2)$,
  where $r_i=X_{i t}^{(i+1)t -1}$ with $t=\frac{n}{C}$. We can consider a three
  dimensional lattice whose $i$-th axis is labelled by all possible values of
  $r_i$. In this space, there is a simple interpretation for a set $T$ to be a
  trapping set, namely, for any $(r_0,r_1,r_2)\in T$ there are three other
  elements $(r'_0,r_1,r_2)$, $(r_0,r'_1,r_2)$ and $(r_0,r_1,r'_2)$ in $T$ that
  can be reached from $(r_0,r_1,r_2)$ by moving along exactly one axis. Notice
  that in this case each hash is equivalent to projecting $(r_0,r_1,r_2)$ onto
  two dimensional planes spanned by different coordinates, for example,
  $h_1(r_0,r_1,r_2)=(r_0,r_1)$ is a projection on the plane spanned by the
  first and second coordinate axes of the lattice. A similar argument holds for
  other values of $C >3$, thus, larger values of $\alpha$. 

  For $C\geq 3$, the set of all $C$-tuples $(r_0,r_1,\dots,r_{C-1})$ is a
  $C$-dimensional lattice. We denote this lattice by $L$. The intersection of
  this lattice by the hyperplane $R_i=r_i$ is a $(C-1)$ dimensional lattice
  defined by
  \begin{align*}
    L (R_i=r_i)=\{&(r_0,\dots,r_{i-1},r_{i+1},\dots,r_{C-1}): \\
      &(r_0,r_1,\dots,r_{i-1},r_i,r_{i+1},\dots,r_{C-1}) \in L\}.
  \end{align*}

  Similarly for $S \subset L$, we have the following definition
  \begin{align*}
    S (R_i=r_i)=\{&(r_0,\dots,r_{i-1},r_{i+1},\dots,r_{C-1}): \\
      &(r_0,r_1,\dots,r_{i-1},r_i,r_{i+1},\dots,r_{C-1}) \in S\}.
  \end{align*}
  Obviously, $S(R_i=r_i) \subset L(R_i=r_i)$. We have the following proposition
  whose proof simply follows from the definition of the trapping set.
  \begin{proposition}\label{intersect_trap}
    Assume that $T$ is a trapping set for the $C$ dimensional lattice
    representation $L$ of the non-zero spectral domain variables as explained
    before. Then for any $r_i$ on the $i$-th axis, $T(R_i=r_i)$ is either empty
    or a trapping set for the $(C-1)$ dimensional lattice $L(R_i=r_i)$.
  \end{proposition}

  \begin{proposition}\label{trap_set_size}
    The size of the trapping set for a $C$ dimensional lattice is at least $2^C$. 
  \end{proposition}

  \begin{IEEEproof}
    We use a simple proof using the induction on $C$. For $C=1$, we have a one
    dimensional lattice along a line labelled with $r_0$. In this case, there
    must be at least two variables on the line to build a trapping set.
    Consider a trapping set $T$ of dimension $C$.  There are at least two
    points $(r_0,r_1,\dots,r_{C-1})$ and $(r'_0,r_1,\dots,r_{C-1})$ in $T$. By
    Proposition \ref{intersect_trap}, $T(R_0=r_0)$ and $T(R_0=r'_0)$ are two
    $(C-1)$ dimensional trapping sets each consisting of at least $2^{C-1}$
    elements by induction hypothesis. Thus, $T$ has at least $2^C$ elements. 
  \end{IEEEproof}

  \begin{remark}
    The bound $|T|\geq2^C$ on the size of the trapping set is actually tight.
    For example, for $i \in [C]$ consider $r_i, r'_i$ where $r_i\neq r'_i$ and
    let  
    $$T=\{(a_0,a_1,\dots,a_{C-1}): a_i \in \{r_i,r'_i\} , i \in [C]\}.$$ 
    It is easy to see that $T$ is a trapping set with $2^C$ elements
    corresponding to the vertices of a $C$ dimensional cube. 
  \end{remark}

  We now prove the following proposition which implies that if the peeling
  decoder can decode all of the variable nodes except a fixed number of them,
  with high probability it can continue to decode all of them.

  \begin{proposition}
    Let $s$ be a fixed positive integer. Assume that $\alpha=1-\frac{1}{C}$ for
    some integer $C \geq 3$ and consider a hash structure with $C$ different
    hashes as explained before. If the peeling decoder decodes all except a set
    of variables of size $s$, it can decode all of the variables with very high
    probability.
  \end{proposition} 

  \begin{IEEEproof}
    The proof in very similar to \cite{Pawar:2013vm}. Let $T$ be a trapping set
    of size $s$. By Proposition \ref{trap_set_size}, we have $s\geq 2^C$. Let
    $p_i$ be the number of distinct values taken by elements of $T$ along the
    $R_i$ axis and let $p_{\max}=\max_{i \in [C]} p_i$. Without loss of
    generality, let us assume that the $R_0$ axis is the one having the maximum
    $p_i$. Consider $T(R_0=r_0)$ for those $p_{\max}$ values of $r_0$ along the
    $R_0$ axis. Proposition \ref{intersect_trap} implies that each $T(R_0=r_0)$
    is a trapping set which has at least $2^{C-1}$ elements according to
    Proposition \ref{trap_set_size}. This implies that $s \geq 2^{C-1}
    p_{\max}$ or $p_{\max} \leq \frac{s}{2^{C-1}}$. Moreover, $T$ being the
    trapping set implies that there are subsets $T_i$ consisting of elements
    from axes $R_i$ and all of the elements of $T$ are restricted to take their
    $i$-th coordinate values along $R_i$ from the set $T_i$. Considering the
    way that we generate the position of non-zero variables $X_0^{n-1}$ with
    the equivalent representation $(r_0,r_1,\dots,r_{C-1})$, the coordinate of
    any variable is selected uniformly and completely independently from on
    another and from the coordinates of the other variables.  This implies that 
    \begin{align*}
      \prob{F_s} &\leq \prob{\text{For any variables in $T$, $r_i \in T_i, i \in [C]$} }\\
      & \leq \prod _{i=0}^{C-1} {{\cal P}i \choose p_i} (\frac{p_i}{{\cal P}_i})^s \leq \prod _{i=0}^{C-1} {{\cal P}i \choose s/2^{C-1}} (\frac{s}{2^{C-1}{\cal P}_i})^s,
    \end{align*}
    where $F_s$ is the event that the peeling decoder fails to decode a
    specific subset of variables of size $s$ and where ${\cal P}_i$ denotes the
    number of all possible values for the $i$-th coordinate of a variable. By
    our construction all ${\cal P}_i$ are equal to
    $P=2^{n/C}=2^{n(1-\alpha)}=N^{(1-\alpha)}$, thus we obtain that
    \begin{align*}
      \prob{F_s} &\leq {P \choose s/2^{C-1}}^C \left(\frac{s}{2^{C-1}P}\right)^{s C}\\
      &\leq \left(\frac{2^{C-1} P e}{s}\right)^{s C/2^{C-1}} \left(\frac{s}{2^{C-1}P}\right)^{s C}\\
      &\leq \left(\frac{s e^{1/(2^{C-1}-1)}}{2^{C-1}P}\right)^{s C(1-1/2^{C-1})}.
    \end{align*}
    Taking the union bound over all ${K \choose s}$ possible ways of selection
    of $s$ variables out of $K$ variables, we obtain that 
    \begin{align*}
      \prob{F} &\leq {K \choose s} \prob{F_s}\\
      & \leq \left(\frac{e P^{C-1}}{s}\right)^s  \left(\frac{s e^{1/(2^{C-1}-1)}}{2^{C-1}P}\right)^{s C(1-1/2^{C-1})}\\
      &=O(1/P^{s(1-\frac{C}{2^{C-1})}}) \\
      &\leq O(1/P^{(2^C- 2C)}) = O(1/N^{\frac{2^C}{C}-2}).
    \end{align*}
    For $C \geq 3$, this gives an upper bound of $O(N^{-\frac{2}{3}})$.
  \end{IEEEproof}

  \subsection{Generalized Hash Construction}
  \slabel{generalized_hash}

  The hash construction that we explained only covers values of
  $\alpha=1-\frac{1}{C}$ for $C\geq 3$ which belongs to the region $\alpha \in
  [\frac{2}{3},1)$. We will explain a hash construction that extends to any
  value of $C$ and $\alpha \in (0,1)$, which is not necessarily of
  the form $1-\frac{1}{C}$. This construction reduces to the very and less sparse
  regimes hash constructions when $\alpha = \frac{1}{C}$, $\alpha\in(0,1/3]$, and
  $\alpha = 1-\frac{1}{C}$, $\alpha\in[2/3,1)$, respectively.

  In the very sparse regime $\alpha=\frac{1}{3}$, we have $C=3$ different
  hashes and for a non-zero spectral variable $X$ with index
  $X_0^{n-1}=(r_0,r_1,r_2)$, $h_i(X_0^{n-1})=r_i$ thus the output of different
  hashes depend on non overlapping parts of the binary index of $X$ whereas for
  $\alpha=\frac{2}{3}$ the hash outputs are $(r_0,r_1)$, $(r_1,r_2)$ and
  $(r_0,r_2)$ which overlap on a portion of binary indices of length
  $\frac{n}{3}$. Intuitively, it is clear that in order to construct different
  hashes for $\alpha \in (\frac{1}{3}, \frac{2}{3})$, we should start
  increasing the overlapping size of different hashes from $0$ for
  $\alpha=\frac{1}{3}$ to $\frac{n}{3}$ for $\alpha=\frac{2}{3}$. We give the
  following construction for the hash functions 
  \begin{align*}
    h_i(X_0^{n-1})=X_{i\, t}^{i\, t + b}, i\in [C],
  \end{align*} 
  where $t=\frac{n}{C}$ and the values of the indices are computed modulo $n$,
  for example $X_n=X_0$. In the terminology of \sref{hashing}, we pick $\Hash_i
  = \Psi_b^T \Sigma_i^T \in\F_2^{k\times n}$,  where $\Sigma_i\in\F_2^{n\times
    n}$ is the identity matrix with columns circularly shifted by $(i+1)b$ to
  the left.  It is clear that each hash is a surjective map from $\F_2^n$ into
  $\F_2^{n \alpha}$. Therefore, if we pick $b=n\alpha$, the number of output
  bins in each hash is $B=2^{n \alpha}=N^\alpha=K$, thus the average number of
  non-zero variables per bin in every hash is equal to $\beta=\frac{K}{B}=1$.
  In terms of decoding performance for the intermediate values of
  $\alpha\in(\frac{1}{3},\frac{2}{3})$, one expects that the performance of the
  peeling decoder for this regime is between the very sparse regime
  $\alpha=\frac{1}{3}$ and the less sparse one $\alpha=\frac{2}{3}$. 

  \section{Experimental Results}
  \slabel{empirical}

  In this section, we empirically evaluate the performance of the SparseFHT
  algorithm for a variety of design parameters.  The simulations are
  implemented in C programming language and the success probability of the
  algorithm has been estimated via sufficient number of trials. We also
  provide a comparison of the run time of our algorithm and the standard
  Hadamard transform. 

  \begin{figure}[t]
    \centering
    \includegraphics[width=\linewidth]{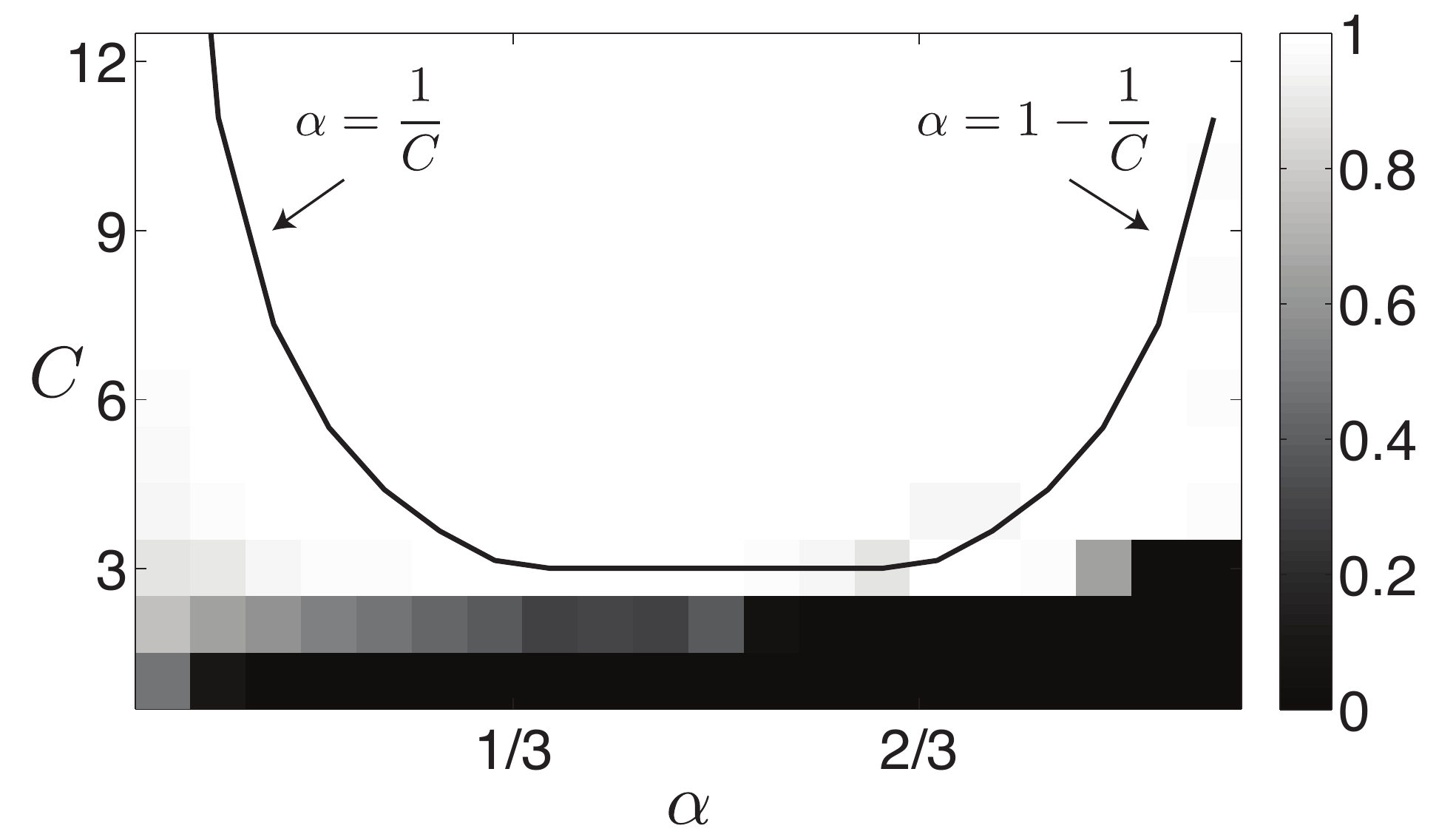}
    \caption{Probability of success of the algorithm in the very sparse regime
      as a function of $\alpha$ and $C$. The dimension of the signal is $N=2^{22}$. 
      The black line corresponds to $\alpha=\frac{1}{C}$ and $\alpha=1-\frac{1}{C}$
      in the very and less sparse regimes, respectively. We fix $\beta=1$. The
      hashing matrices are deterministically picked as described in
      \sref{generalized_hash}.
    }
    \flabel{very_sparse_success}
  \end{figure}

  \begin{figure}[t]
    \centering
    \includegraphics[width=\linewidth]{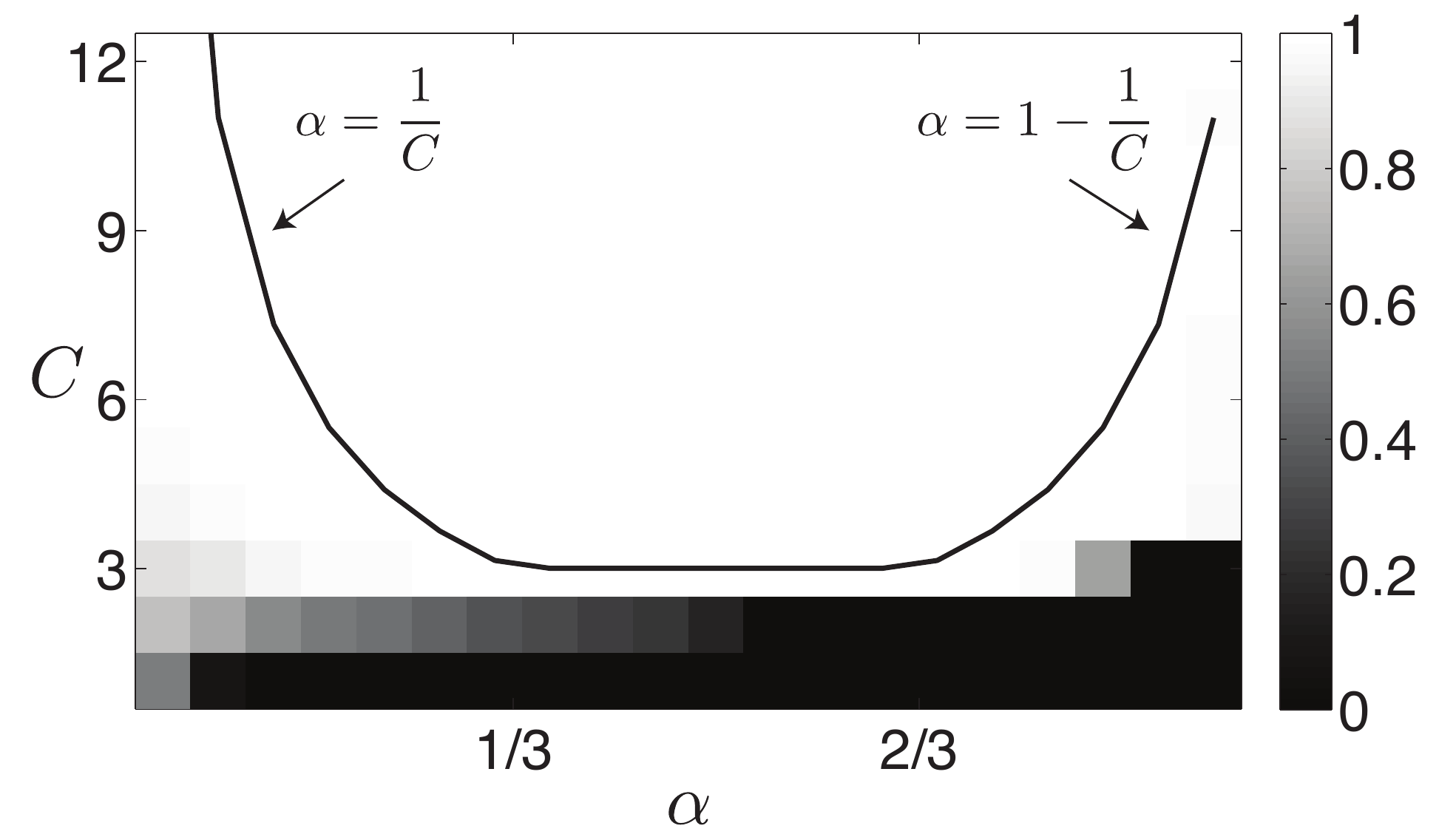}
    \caption{Probability of success of the algorithm in the very sparse regime
      as a function of $\alpha$ and $C$. The dimension of the signal is
      $N=2^{22}$.  The black line corresponds to $\alpha=\frac{1}{C}$ and
      $\alpha=1-\frac{1}{C}$ in the very and less sparse regimes, respectively.
      We fix $\beta=1$. The hashing matrices are picked at random for every
      trial.
    }
    \flabel{very_sparse_success_random}
  \end{figure}

  \begin{figure}[t]
    \centering
    \includegraphics[width=\linewidth]{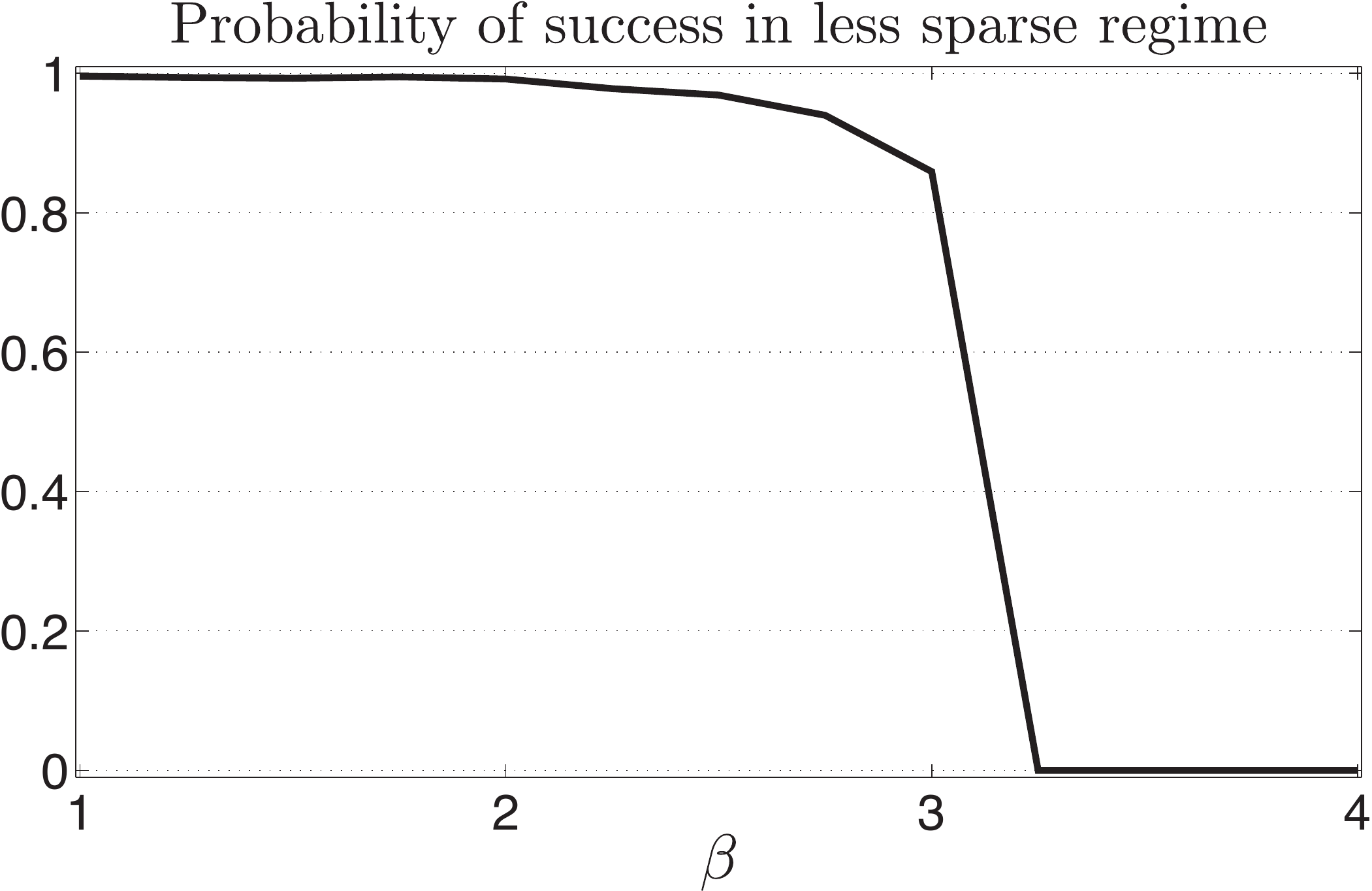}
    \caption{Probability of success of the algorithm in the less sparse regime
      as a function of $\beta=K/B$. We fix $N=2^{22}$, $B=2^{17}$, $C=4$, and
      vary $\alpha$ in the range $0.7$ to $0.9$.
    }
    \flabel{thresh_behav}
  \end{figure}

  \begin{itemize} 
    \item \textit{Experiment 1}: We fix the signal size to $N=2^{22}$ and run
      the algorithm $1000$ times to estimate the success probability for
      $\alpha \in (0, \frac{1}{3}]$ and $1\leq C\leq 12$. The hashing
      scheme used is as described in \sref{generalized_hash}.
      \ffref{very_sparse_success} shows the simulation result.  Albeit the
      asymptotic behavior of the error probability is only guaranteed for
      $C = (\frac{1}{\alpha} \vee \frac{1}{1-\alpha})$, we observe much
      better results in practice. Indeed, $C=4$ already gives a probability
      of success very close to one over a large range of $\alpha$, and only
      up to $C=6$ seems to be required for the largest values of $\alpha$.

    \item \textit{Experiment 2}: We repeat here experiment 1, but instead of
      deterministic hashing matrices, we now pick $\Sigma_i$, $i\in[C]$,
      uniformly at random from $\GL$. The result is shown in
      \ffref{very_sparse_success_random}.  We observer that this scheme
      performs at least as well as the deterministic one.

    \item \textit{Experiment 3}: In this experiment, we investigate the
      sensitivity of the algorithm to the value of the parameter
      $\beta=K/B$; the average number of non-zero coefficients per bin. As
      we explained, in our hash design we use $\beta \approx 1$. However,
      using larger values of $\beta$ is appealing from a computational
      complexity point of view.  For the simulation, we fix $N=2^{22}$,
      $B=2^{17}$, $C=4$, and vary $\alpha$ between $0.7$ and $0.9$, thus
      changing $K$ and as a result $\beta$.  \ffref{thresh_behav} show the
      simulation results.  For $\beta\approx 0.324$, the algorithm succeeds
      with probability very close to one. Moreover, for values of $\beta$
      larger than $3$, success probability sharply goes to $0$.

    \item \textit{Runtime measurement}: We compare the runtime of the SparseFHT
      algorithm with a straightforward implementation of the fast Hadamard
      transform. The result is shown in \ffref{runtime}  for $N=2^{15}$.
      SparseFHT performs much faster for $0<\alpha < 2/3$.

      It is also intersting to identify the range of $\alpha$ for which
      SparseFHT has a lower runtime than the conventional FHT. We define $\alpha^*$,
      the largest value of $\alpha$ such that SparseFHT is faster than FHT for any
      lower value of $\alpha$. That is
      \begin{align*}
        \alpha^* = \sup_{\alpha\in(0,1)}\{\alpha\,:\, \forall \alpha^\prime \leq \alpha,\ T_{FHT}(n) > T_{SFHT}(\alpha^\prime,n)\},
      \end{align*}
      where $T_{FHT}$ and $T_{SFHT}$ are the runtimes of the conventional FHT and SparseFHT, respectively.
      We plot $\alpha^*$ as a function of $n=\log_2N$ in \ffref{improv_zone}.
      
  \end{itemize}

  \begin{figure}
    \centering
    \includegraphics[width=\linewidth]{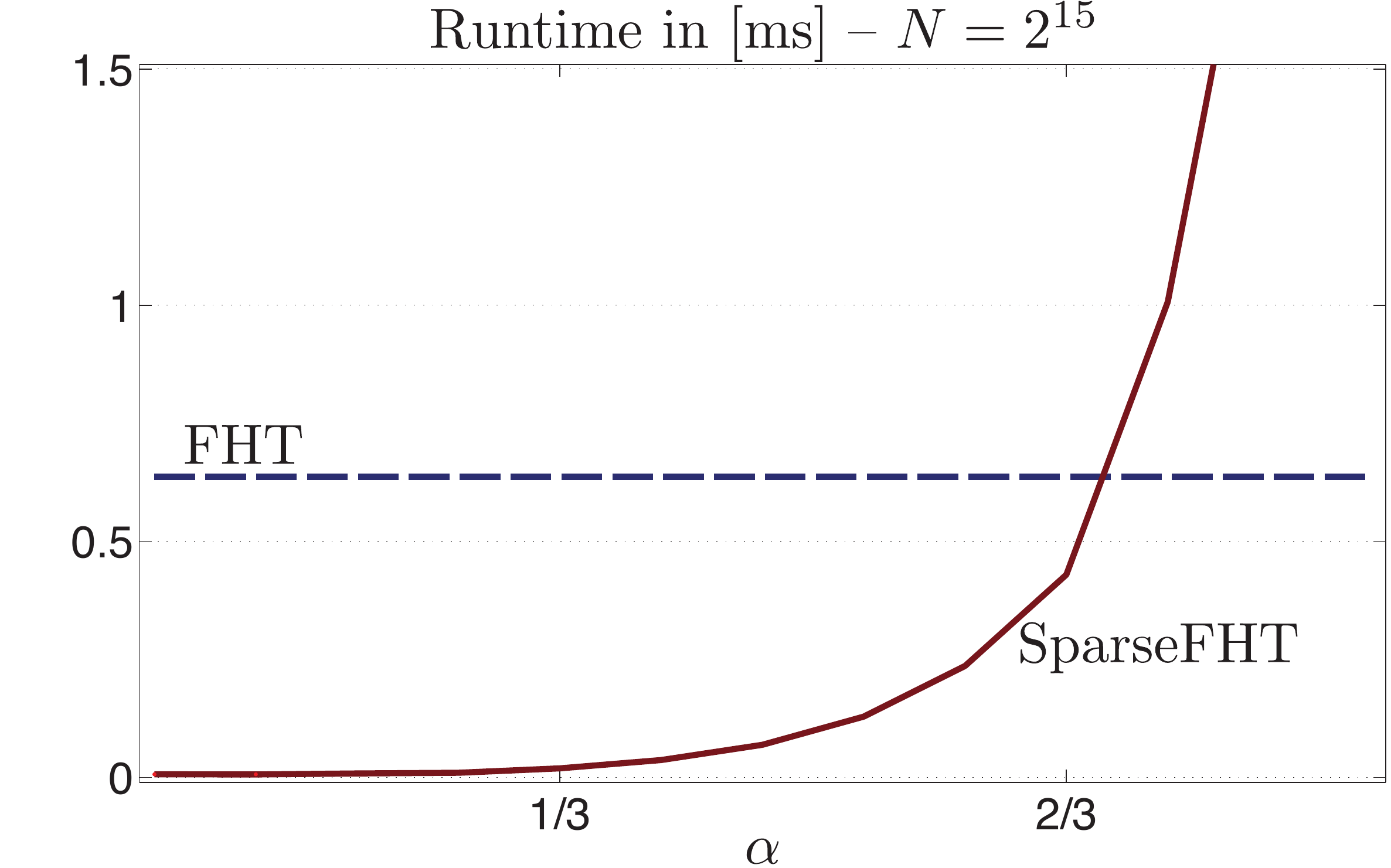}
    \caption{Comparison of the Median runtime in ms of the SparseFHT and
      conventional FHT for $N=2^{15}$ and for different values of $\alpha$.
      Confidence interval where found to be negligible and are omitted here.
      Lower runtime is better.}
    \flabel{runtime}
  \end{figure}

  \begin{figure}
    \centering
    \includegraphics[width=\linewidth]{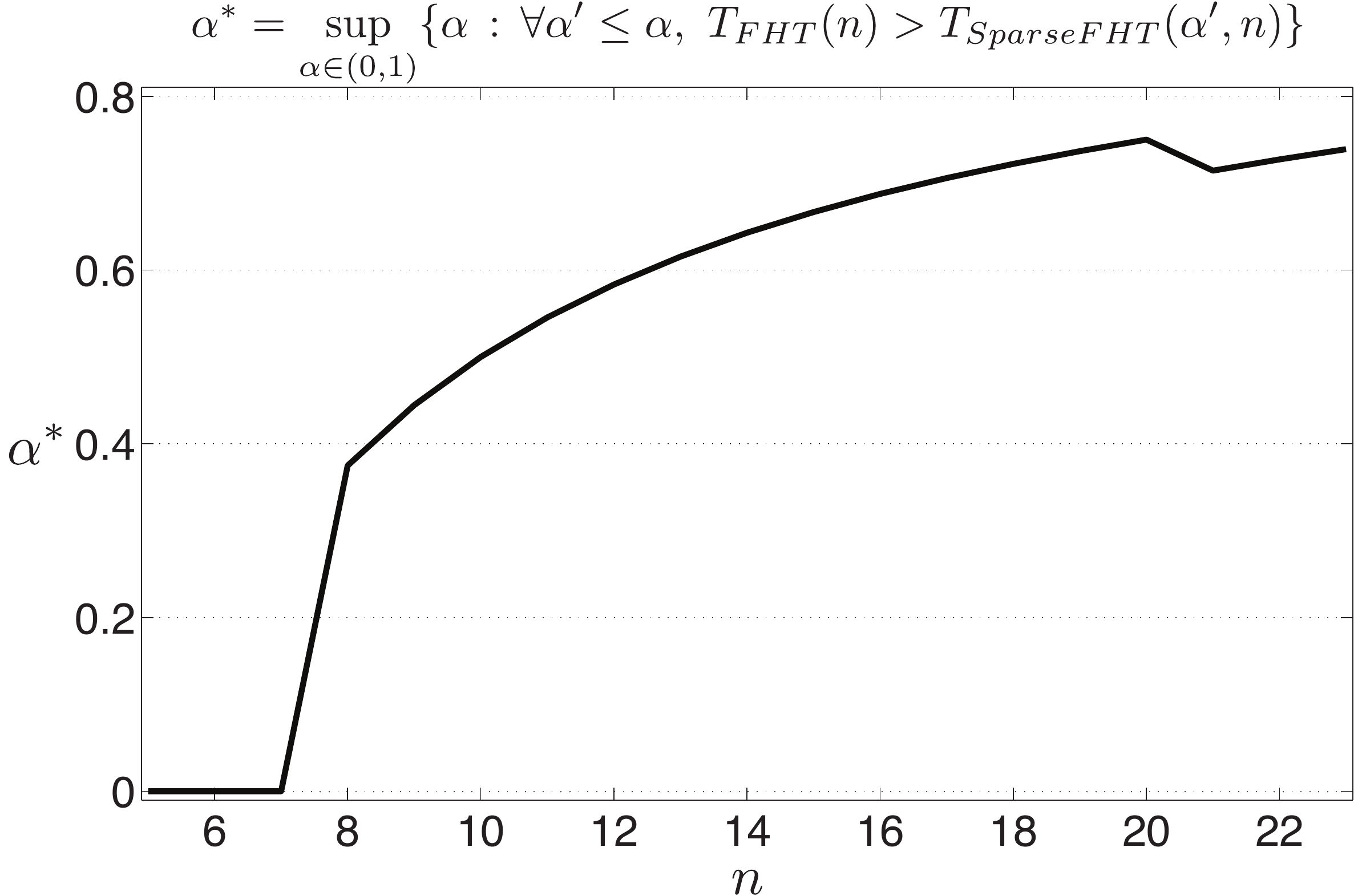}
    \caption{In this figure, we plot $n=\log_2N$ against $\alpha^*$, the largest value of $\alpha$ such that
      SparseFHT runs faster than the conventional FHT for all values of $\alpha$ smaller or equal.
      When FHT is always faster, we simply set $\alpha^*=0$. Larger values are better.}
    \flabel{improv_zone}
  \end{figure}

  \begin{remark}
    In the computation of the complexity in \sref{complexity_analysis}, we have
    assumed that matrix-vector multiplications in $\F_2^n$ can be done in
    $\bigO(1)$.  In general, it is not true. However, the deterministic
    hashing scheme of the algorithm is nothing but a circular bit shift that
    can be implemented in a constant number of operations, independent of the
    vector size $n$.

    If one is given $\Sigma$, some matrix from $\F_2^{n\times n}$, and its
    inverse transpose $\Sigma^{-T}$, the overall complexity of the algorithm
    would nonetheless be unchanged. First, we observe that it is possible to
    compute the inner product of two vectors in constant time using bitwise
    operations and a small look-up
    table\footnote{\url{http://graphics.stanford.edu/~seander/bithacks.html#ParityLookupTable}}.
    Now, given the structure of $\Psi_b$, computing $\Sigma\Psi_b m$ in
    \algref{fasthadhash} only requires $\log_2K$ inner products. Thus the
    complexity of \algref{fasthadhash} is unchanged.  Finally, \eref{index_rec}
    can be split into pre-computing $\Sigma^{-T}\Psi_b k$ at the same time as
    we subsample the signal (in $\bigO(\log_2K)$), and computing the inner
    product between $\hat{v}$ and the $n-b$ first columns of $\Sigma$ when doing
    the decoding ($\bigO(\log_2\frac{N}{K})$).
  \end{remark}

  \section{Conclusion}

  We presented a new algorithm to compute the Hadamard transform of a signal
  of length $N$ which is $K$-sparse in the Hadamard domain. The algorithm
  presented has complexity $\bigO(K\log_2 K \log_2\frac{N}{K})$ and only
  requires $\bigO(K\log_2\frac{N}{K})$ time-domain samples. We show that
  the algorithm correctly reconstructs the Hadamard transform of the signal
  with high probability asymptotically going to one. 
  
  The performance of the algorithm is also evaluated empirically through
  simulation, and its speed is compared to that of the conventional fast
  Hadamard transform. We find that considerable speed-up can be obtained, even
  for moderate signal length (e.g. $N=2^{10}$) with reasonnable sparsity
  assumptions.

  However, from the statement of \pref{collision_detect}, it will be apparent
  to the reader that the algorithm is absolutely not robust to noise. In fact,
  at very large signal size, the machine noise, using double-precision floating
  point arithmetic, proved to be problematic in the simulation. To make the
  algorithm fully practical, a robust estimator is needed to replace
  \pref{collision_detect}, and is, so far, left for future work.

  \bibliographystyle{IEEEtran} 
  \bibliography{bibliography}

  \appendices

  \section{Proof of the Properties of the WHT}
  \slabel{prop_proof}

  \subsection{Proof of Property~\ref{had_shift}}

  \begin{equation}
    \sum_{m\in\F_2^n} (-1)^{\ip{k}{m}} x_{m+p} = \sum_{m\in\F_2^n} (-1)^{\ip{k}{m+p}} x_m.
    \nonumber
  \end{equation}
  And the proof follows by taking $(-1)^{\ip{k}{p}}$ out of the sum and
  recognizing the Hadamard transform of $x_m$. \hfill $\blacksquare$

  \subsection{Proof of Property \ref{gln2perm}}

  As we explained, it is possible to assign an $N \times N$ matrix $\Pi$ to the
  permutation $\pi$ as follows
  \begin{equation}
    (\Pi)_{i,j} = \begin{cases}
      1 & \text{if $j=\pi(i) \Leftrightarrow i=\pi^{-1}(j)$} \\
      0 & \text{otherwise.}
    \end{cases}.
    \nonumber
  \end{equation} 
  Let $\pi_1$ and $\pi_2$ be the permutations associated with $\Pi_1$ and
  $\Pi_2$.  Since $(H_N)_{i,j} = (-1)^{\ip{i}{j}}$, the identity
  \eqref{perm_prop} implies that 
  \begin{equation}
    (-1)^{\ip{\pi_2(i)}{j}} = (-1)^{\ip{i}{\pi_1^{-1}(j)}}.
    \nonumber
  \end{equation}
  Therefore, for any $i,j \in \F_2^n$, $\pi_1,\pi_2$ must satisfy
  $\ip{\pi_2(i)}{j} = \ip{i}{\pi_1^{-1}(j)}$.  By  linearity of the inner
  product, one obtains that
  \begin{eqnarray}
    \ip{\pi_2(i+k)}{j} & = & \ip{i+k}{\pi_1^{-1}(j)} \nonumber \\
    & = & \ip{i}{\pi_1^{-1}(j)} + \ip{k}{\pi_1^{-1}(j)} \nonumber \\
    & = & \ip{\pi_2(i)}{j} + \ip{\pi_2(k)}{j}. \nonumber
  \end{eqnarray}
  As $i,j \in \F_2^n$ are arbitrary, this implies that $\pi_2$, and by symmetry
  $\pi_1$, are both linear operators. Hence, all the permutations satisfying
  \eqref{perm_prop} are in one-to-one correspondence with the elements of
  $\GL$. \hfill $\blacksquare$

  \subsection{Proof of Property~\ref{had_perm}}

  Since $\Sigma$ is non-singular, then $\Sigma^{-1}$ exists. It follows from
  the definition of the WHT that
  \begin{align*}
    \sum_{m\in\F_2^n} (-1)^{\ip{k}{m}}x_{\Sigma m}  &= \sum_{m\in\F_2^n} (-1)^{\ip{k}{\Sigma^{-1} m}} x_m \\
    &= \sum_{m\in\F_2^n} (-1)^{\ip{\Sigma^{-T}k}{m}} x_m. \tag*{\qedhere}
  \end{align*}
  This completes the proof. \hfill $\blacksquare$

  \subsection{Proof of Property~\ref{had_ds}}

  \begin{multline}
    \sum_{m\in\F_2^b} (-1)^{\ip{k}{m}}x_{\Psi_b m}  \nonumber \\
    \hfill \begin{array}{cl}
      = & \frac{1}{\sqrt{N}} \sum\limits_{m\in\F_2^b}(-1)^{\ip{k}{m}}\sum\limits_{p\in\F_2^n}(-1)^{\ip{\Psi_bm}{p}}X_p \nonumber \\
      = &  \frac{1}{\sqrt{N}} \sum\limits_{p\in\F_2^n} X_p \sum\limits_{m\in\F_2^b} (-1)^{\ip{m}{k + \Psi_b^Tp}}.  \nonumber
    \end{array}
  \end{multline}
  In the last expression, if $p=\Psi_bk+ i$ with $i\in\Null(\Psi_b^T)$ then it
  is easy to check that the inner sum is equal to $B$, otherwise it is equal to
  zero. Thus, by proper renormalization of the sums one obtains the proof.
  \hfill $\blacksquare$

  \section{Proof of \pref{collision_detect}}
  \slabel{sup_est_proof}

  We first show that if multiple coefficients fall in the same bin, it is very
  unlikely that 1) is fulfilled.  Let $\mathcal{I}_k =
  \{j\,|\Hash j=k\}$ be the set of variable indices hashed to bin $k$.
  This set is finite and its element can be enumerated as $\mathcal{I}_k = \{j_1,\ldots,j_{\frac{N}{B}}\}$.
  We show that a set $\{X_j\}_{j\in\mathcal{I}_k}$ is very unlikely, unless it
  contains only one non-zero element. Without loss of generality, we consider
  $\sum_{j\in\mathcal{I}_k} X_j=1$. Such $\{X_j\}_{j\in\mathcal{I}_k}$ is a
  solution of
  \begin{equation}
    \begin{bmatrix}
      1 & \cdots & 1 \\
      (-1)^{\ip{\sigma_1}{j_1}} & \cdots & (-1)^{\ip{\sigma_1}{j_{\frac{N}{B}}}} \\
      \vdots & \ddots & \vdots \\
      (-1)^{\ip{\sigma_{n-b}}{j_1}} & \cdots & (-1)^{\ip{\sigma_{n-b}}{j_{\frac{N}{B}}}}
    \end{bmatrix}
    \begin{bmatrix}
      X_{j_1} \\ \vdots \\ X_{j_{\frac{N}{B}}}
    \end{bmatrix}
    =
    \begin{bmatrix}
      1 \\ \pm 1 \\ \vdots \\ \pm 1
    \end{bmatrix},
    \nonumber
  \end{equation}
  where $\sigma_i, i\in\{1,\ldots,n\}$ denotes the $i$-th column of the matrix $\Sigma$.
  The left hand side matrix in the expression above, is $(n-b+1)\times
  2^{n-b}$.  As $\sigma_1,\ldots,\sigma_{n-b}$ form a basis for
  $\mathcal{I}_k$,  all the columns are different and are (omitting the top
  row) the exhaustive list of all $2^{n-b}$ possible $\pm 1$ vectors.  Thus the
  right vector is always one of the columns of the matrix and there is a
  solution with only one non-zero component ($1$-sparse solution) to this
  system whose support can be uniquely identified. Adding any vector from the
  null space of the matrix to this initial solution yields another solution.
  However, as we will show, due to its structure this matrix is full rank and
  thus its null space has dimension $2^{n-b}-n+b-1$. Assuming a continuous
  distribution on the non-zero components $X_i$, the probability that
  $\{X_i\}_{i\in\mathcal{I}_k}$ falls in this null space is zero.

  To prove that the matrix is indeed full rank, let us first focus on the rank
  of the sub-matrix obtained by removing the first row. This submatrix itself
  always contains $M = -2 I + \mathds{1} \mathds{1}^T$, where $I$ is the
  identity matrix of order $n-b$ and $\mathds{1}$ is the all-one vector of
  dimension $(n-b)$.  One can simply check that $M$ is a symmetric matrix, thus
  by spectral decomposition, it has $n-b$ orthogonal eigen-vectors $v_i, i\in
  [n-b]$.  It is also easy to see that the normalized all-one vector
  $v_0=\frac{\mathds{1}}{\sqrt{n-b}} $ of dimension $n-b$ is an eigen-vector of
  $M$ with eigen-value $\lambda_0=n-b-2$. Moreover, assuming the orthonormality
  of the eigen-vectors, it results that $v_i^T M v_i= \lambda_i = -2$, where we
  used $v_i^T \mathds{1}=v_i^T v_0=0$ for $i \neq 0$. Thus, for $n-b \neq 2$
  all of the eigen-vlaues are non-zero and $M$ is invertible, which implies
  that the sub-matrix resulted after removing the first row is full rank.
  In the case where $n-b=2$, one can notice that the Hadamard matrix of size 2
  will be contained as a submatrix, and thus the matrix will be full rank.


  Now it remains to prove that initial matrix is also full rank with a rank of
  $n-b+1$. Assume that the columns of the matrix are arranged in the
  lexicographical order such that neglecting the first row, the first and the
  last column are  all $1$ and all $-1$.  If we consider any linear combination
  of the rows except the first one, it is easy to see that the first and the
  last element in the resulting row vector have identical magnitudes but
  opposite signs. This implies that the all-one row cannot be written as a
  linear combination of the other rows of the matrix.  Therefore, the rank of
  the matrix must be $n-b+1$.

  To prove \eref{index_rec}, let $\Sigma_L$ and $\Sigma_R$ be the matrices
  containing respectively the first $n-b$ and the last $b$ columns of $\Sigma$,
  such that $\Sigma = [\Sigma_L\,\Sigma_R]$. If there is only one coefficient
  in the bin, then \eref{ip_est} implies  that $\hat{v} = [\, (j^T\Sigma_{L}) \
  0 \,]^T$.  Using definitions \eref{ds_mat} and \eref{hashing}, we obtain that
  $\Psi_b\Hash j = [\,0 \ (j^T\Sigma_R)\,]^T$.  We observe that they
  sum to $\Sigma^T j$ and the proof follows.  \hfill $\blacksquare$

  \section{Proof of Proposition \ref{randomsupp}}\label{proof_randomsupp}
  
  For $t \in [K]$, let $H_t$ denote the size of the random set obtained by
  picking $t$ objects from $[N]$ independently and uniformly at random with
  replacement. Let $a_t$ and $v_t$ denote the average and the variance of $H_t$
  for $t \in [K]$. It is easy to see that $\{H_t\}_{t \in [K]}$ is a Markov
  process. Thus, we have
  \begin{align*}
    \E{H_{t+1} -H_t |H_t}=(1-H_t / N),
  \end{align*}
  because the size of the random set increases if an only if we choose an
  element from $[N]\backslash H_t$. This implies that $a_{t+1}=1+\gamma a_t$,
  where $\gamma=1-\frac{1}{N}$. Solving this equation we obtain that 
  \begin{align}\label{at_form}
    a_t=\sum_{r=0}^t \gamma ^r=\frac{1-\gamma ^{t+1}}{1-\gamma}= N(1-\gamma^{t+1}).
  \end{align}
  In particular, $a_K=N(1-(1-\frac{1}{N})^K)$, which implies that
  $\E{\frac{H_K}{K}}=\frac{N}{K} (1-(1-\frac{1}{N})^K)$. One can check that for
  $K=N^\alpha$, $0<\alpha <1$, as $N$ tends to infinity $\E{\frac{H_K}{K}}$
  converges to $1$. To find the variance of $H_t$, we use the formula 
  \begin{align}\label{varform}
    \var(H_{t+1})=\var(H_{t+1}|H_t)+\var(\E{H_{t+1}|H_t)}).
  \end{align}
  Therefore, we obtain that 
  \begin{align}\label{var1}
    \var(\E{H_{t+1}|H_t})=\var(1+\gamma H_t)=\gamma^2 v_t.
  \end{align}
  Moreover, for the first part in \eqref{varform}, we have
  \begin{align}\label{var2}
    \var(H_{t+1}|H_t)&=\mathbb{E}_{H_t}\{\var(H_{t+1}|H_t=h_t)\}\nonumber\\
    &=\mathbb{E}_{H_t}\{\var(H_{t+1}-H_t|H_t=h_t)\}\nonumber\\
    &\stackrel{(I)}{=}\E{\frac{H_t}{N} \left(1-\frac{H_t}{N}\right)}\nonumber\\
    &= \frac{a_t}{N} + \frac{a_t^2+v_t}{N^2},
  \end{align}
  where in $(I)$ we used the fact that given $H_t$, $H_{t+1}-H_t$ is a
  Bernoulli random variable with probability $\frac{H_t}{N}$, thus its variance
  in equal to $\frac{H_t}{N}(1-\frac{H_t}{N})$. Combining \eqref{var1} and
  \eqref{var2}, we obtain that 
  \begin{align}\label{vareq}
    v_{t+1}=\left(\gamma^2 + \frac{1}{N^2}\right) v_t + \frac{a_t}{N}\left(1+\frac{a_t}{N}\right). 
  \end{align}
  From \eqref{at_form}, it is easy to see that $a_t$ is increasing in $t$.
  Moreover, from \eqref{vareq} it is seen that $v_{t+1}$ is increasing function
  of $a_t$, thus, if we consider the following recursion 
  \begin{align*}
    w_{t+1}=\left(\gamma^2+\frac{1}{N^2}\right) w_t + \frac{a_K}{N}\left(1+\frac{a_K}{N}l\right),
  \end{align*}
  then for any $t\in[K]$, $v_t \leq w_t$. As $w_t$ is also an increasing
  sequence of $t$, we obtain that 
  \begin{align*}
    v_K &\leq w_K \leq w_\infty = \frac{a_K}{N}\left(1+\frac{a_K}{N}\right) / \left(1-\gamma^2 - \frac{1}{N^2}\right)\\
    &=\frac{a_K}{2} \left(1+\frac{a_K}{N}\right)/\left(1-\frac{1}{N}\right).
  \end{align*}
  Using Chebyshev's inequality, we obtain that for any $\epsilon>0$
  \begin{align*}
    \prob{\frac{H_K}{K} \geq (1+\epsilon)} \leq \frac{v_K }{K^2 (\epsilon + 1-\frac{a_K}{K})^2 } = \Theta\left(\frac{1}{\epsilon^2 K}\right).
  \end{align*}
  Obviously, $\frac{H_K}{K} \leq 1$, thus $\frac{H_K}{K}$ converges to $1$ in
  probability as $N$ and as a result $K$ tend to infinity. \hfill $\blacksquare$

  \section{Proof of Proposition \ref{expander}}\label{expander_arg}

  Let $S$ be any set of variable nodes of size at most $\eta K$, where we will
  choose $\eta$ later. 
  the average degree of variable nodes in $S$ is $C$.  Let $\Null_i(S), i \in
  [C]$ be the check neighbors of $\G$ in hash $i$. If for at least one of the
  hashes $i \in [C]$, $|\Null_i(S)| > \frac{|S|}{2}$, it results that there is
  at least one check node of degree $1$ ( a singleton) among the neighbors,
  which implies that the peeling decoder can still proceed to decode further
  variable nodes. 

  Let ${\cal E}^i_s$ denote the event that a specific subset $A$ of size $s$ of
  variable nodes has at most $\frac{s}{2}$ check neighbors in hash $i$. Also
  let ${\cal E}_s=\cap _{i=1}^C {\cal E}^i_s$. By the construction of $\G$, it
  is easy to see that $\prob{{\cal E}_s}=\prod _{i=1}^C \prob{{\cal E}_s^i}$.
  Let $T$ be any subset of check nodes in hash $i$ of size $\frac{s}{2}$. The
  probability that all the neighbors of $A$ in hash $i$ belong to a specific
  set $T$ of size $\frac{s}{2}$ is equal to $(\frac{s}{2 B})^s$. Taking a union
  bound over ${B \choose s/2}$ of all such sets, it is seen that $\prob{{\cal
      E}_s} \leq {B \choose s/2} (\frac{s}{2 B})^s$, which implies that
  $\prob{{\cal E}^i_s} \leq \big({B \choose s/2} (\frac{s}{2 B})^{s }\big)^C$. Taking a
  union bound over all possible subsets of size $s$ of variables, we obtain
  that 
  \begin{align*}
    \prob{F_s} &\leq {K \choose s} \prob{{\cal E}_s} \leq {K \choose s} \left({B \choose s/2} \left(\frac{s}{2 B}\right)^s \right)^C\\
    &\leq \left(\frac{e K}{s}\right)^s \left(\frac{2 e B}{s}\right)^{s C/2} \left(\frac{s}{2B}\right)^{s C}\leq \frac{u^s s^{s(C/2 -1)}}{K^{s(C/2 -1)}},
  \end{align*}
  where $u=e^{C/2+1} (\frac{\beta}{2})^{C/2}$ and where $F_s$ denotes the event
  that the peeling decoder fail to decode a set of variables of size $s$. We
  also used the fact that for $n\geq m$, ${n \choose m} \leq (\frac{n\,
    e}{m})^m$ and $\prob{F_1}=\prob{F_2}=0$. Selecting $\eta=\frac{1}{2
    u^{2/(C-2)}}$ and applying the union bound, we obtain that
  \begin{align*}
    \prob{F}&\leq \sum _{s=1}^{\eta K} \prob{F_s}= \sum _{s=3}^{\eta K} \prob{F_s}=\sum_{s=3}^{\eta K}  \frac{u^s s^{s(C/2 -1)}}{K^{s(C/2 -1)}}\\
    &=O\left(\frac{1}{K^{3(C/2-1)}}\right) + \sum_{s=4}^{\eta K} \left(\frac{1}{2}\right)^s=O\left(\frac{1}{K^{3(C/2-1)}}\right),
  \end{align*}
  where $F$ is the event that the peeling decoder fails to decode all the
  variables. This completes the proof. \hfill $\blacksquare$

\end{document}